\documentclass[10pt]{article}

\usepackage{main_style}
\usepackage[margin=1in]{geometry}

\newtheorem{theorem}{Theorem}

\newtheorem{lemma}[theorem]{Lemma}

\begin{document}
\maketitle

\begin{abstract}
Leader election is one of the fundamental and well-studied problems in distributed computing.
In this paper, we initiate the study of leader election using mobile agents. Suppose $n$ agents are positioned initially arbitrarily on the nodes of an  arbitrary, anonymous,  $n$-node, $m$-edge graph $G$. 
The agents relocate themselves autonomously on the nodes of $G$ and elect an agent as a leader such that the leader agent knows it is a leader and the other agents know they are not leaders. The objective is to minimize 
time and memory requirements. Following the literature, we consider the synchronous setting in which each agent performs its operations synchronously with others and hence the time complexity can be measured in rounds. The quest in this paper is to provide solutions without agents knowing any graph parameter, such as $n$, a priori. 
We first establish that, without agents knowing any graph parameter a priori, there exists a deterministic algorithm to elect an agent as a leader in $O(m)$ rounds with $O(n\log n)$ bits at each agent. 
Using this leader election result, we develop a deterministic algorithm for agents to construct a minimum spanning tree of $G$ in $O(m+n\log n)$ rounds using $O(n \log n)$ bits memory at each agent, without agents knowing any graph parameter a priori.  Finally, using the same leader election result, we provide improved time/memory results for other fundamental distributed graph problems, namely, gathering, maximal independent set, and minimal dominating sets, removing the assumptions on agents knowing graph parameters a priori. 
\end{abstract}



\section{Introduction}
The well-studied {\em message-passing} distributed computing model assumes an underlying distributed network represented as an undirected graph $G=(V,E)$, where each vertex/node corresponds to a {\em computational device} (such as a computer or a processor), and each edge corresponds to a bi-directional communication link. Each node $v\in G$ has a distinct $\Theta(\log n)$-bit identifier, $n=|V|$. 
The structure of $G$ (topology, latency) is assumed to be not known in advance, and
each node typically knows only its neighboring nodes. 
The nodes interact with one another by sending messages (hence the name {\em message-passing}) to achieve a common goal. 
The computation proceeds according to synchronized {\em rounds}. In each round, each node $v$ can perform unlimited local computation and may send a distinct message to each of its neighbors. Additionally, each node $v$ is assumed to have no limit on storage. 
In the {\sf LOCAL} variant of this model, there is no limit on bandwidth, i.e., a node can send any size message to each of its neighbors. In the {\sf CONGEST} variant, bandwidth is taken into account, i.e.,  a node may send only a, possibly distinct, $O(\log n)$-bit message to each of its neighbors.

\begin{table}[htb]
\footnotesize{
\centering
\begin{tabular}{ccccc}
\toprule
{\bf Model} & {\bf Devices} & {\bf Local} & {\bf Device} & {\bf Neighbor} \\ 
& & {\bf computation} & {\bf storage} & {\bf communication} \\ 
\toprule
Message-passing & Static & Unlimited & No restriction & Messages \\ 
\hline
Agent-based & Mobile & Unlimited & Limited & Relocation \\ 
\bottomrule
\end{tabular}
\caption{Comparison of the message-passing and agent-based models. 
\label{table:model-comparison}
}
}
\vspace{-3mm}
\end{table}

In this paper,  
we consider the {\em agent-based} distributed computing model where the computational devices are modeled as  {\em relocatable or mobile computational devices} (which we call agents). Departing from the notion of vertex/node as a {\em static} device in the message-passing model,  the vertices/nodes serve as {\em containers} for the devices in the agent-based model. 
The agent-based model has two major differences with the message-passing model (Table \ref{table:model-comparison} compares the properties of the two models). 

\begin{itemize}
\item [] {\bf Difference I.} The graph nodes do not have identifiers, computation ability, and storage,  but the devices are assumed to have distinct $O(\log n)$-bit identifiers, computation ability, and (limited) storage. 
\item [] {\bf Difference II.} The devices cannot send messages to other devices except the ones co-located at the same node. To send a message to a device positioned at a neighboring node, a device needs to relocate to the neighbor and can exchange information if a device is positioned at the neighbor. 
\end{itemize}
Difference II is the major problem for the agent-based model. 
To complicate further, while a device relocates to a neighbor, the device at that neighbor might relocate to another neighbor. 
Therefore, the devices need to coordinate 
to achieve the common goal. 

In this paper, we initiate the study of a graph-level task of leader election in a distributed network under the agent-based model.  
Leader election is one of the fundamental and well-studied problems in distributed computing due to its applications in numerous problems, such as resource allocation, reliable replication, load balancing, synchronization, membership maintenance, crash recovery, etc. Leader election can also be seen as a form of symmetry breaking, where exactly one special process or node (say a leader) is allowed to make some critical decisions. 
The problem of leader election in the agent-based model requires a set of agents operating in the distributed network to elect a unique leader among themselves, i.e., exactly one agent must output the decision that it is the leader. 

We develop a deterministic algorithm for leader election with provable guarantees on two performance metrics that are fundamental to the agent-based model: {\em time complexity} of a solution and {\em storage requirement} per agent. We focus on the {\em deterministic} algorithms since they may be more suitable for relocatable devices. Our quest is to provide an algorithm that does not ask the agents to rely on any knowledge (neither exact nor an upper bound) on graph parameters, such as $n$ (the network size and also the number of agents), $\Delta$ (the maximum degree of $G$), and $D$ (diameter of $G$). This is in contrast to the message-passing model which typically assumes that $n$ (exact $n$ or an upper bound $N$ on $n$) is known to the nodes/devices, and may be additionally $\Delta$ and $D$ \cite{ChangPZ19}. This also contrasts research in the agent-based model with known parameters (e.g., \cite{ChandMS23,Molla-2021IPDPS,PattanayakBCM24}).   On the one hand, not knowing these parameters has its own merits as the solutions designed are more resilient to network changes and device faults. On the other hand, algorithm design becomes challenging since devices may not know how long to run a procedure to guarantee a solution. 

Moreover, the agent-based model treats storage requirement as the first order performance metric in addition to time complexity. This is in contrast to the message-passing model where storage complexity was often neglected with the implicit assumption that the devices have no restriction on the amount of storage needed to successfully run the algorithm; in the message-passing model, the focus was given on {\em message complexity} (the total number of messages sent by all nodes for a solution \cite{Pandurangan0S18}) as the first order performance metric in addition to time complexity.   The goal is to use storage as small as possible (comparable to the device identifier size of $O(\log n)$ bits per device). The limited storage makes it impossible for the relocatable devices to first traverse the graph to learn the topology and then run graph computation as a second step.

Using the proposed deterministic leader election algorithm  with provable guarantees on time and storage, we 
construct a minimum spanning tree (MST) of $G$, another fundamental and well-studied problem in distributed computing, for the first time in the agent-based model,  without agents knowing any graph parameter a priori. 
We provide both time and memory complexities. 
Finally, as an application, using the same leader election result, we provide improved time/memory complexity algorithms for many other fundamental distributed graph problems, namely gathering, maximal independent set (MIS), and minimal dominating sets (MDS), removing the parameter assumptions in the literature. 

\subsection{Motivation}
 The agent-based model is useful and applicable in scenarios like private networks in the military or sensor networks in inaccessible terrain where direct access to the network is possibly obstructed, but small battery-powered relocatable computational devices can navigate to learn network structures and their properties for overall network management. Prominent use of agent-based model in network management can be seen in areas such as underwater navigation \cite{Cong2021}, network-centric warfare in military systems \cite{LEE2018}, modeling social network \cite{Zhuge2018}, studying social epidemiology \cite{ELSAYED2012}, etc. Additionally, limited storage may not reveal much about the network even when the devices are compromised. Furthermore, device relocation for communication helps to not worry too much about message compromise.

The agent-based model has recently found its use in various areas. 
One prominent example is Martinkus {\it et al.} \cite{Agentnet} which proposes {\sf AgentNet}  -- a graph neural network (GNN) architecture, in which a collection of (neural) relocatable devices (called neural agents) {\em walk} the graph and collectively classify the graph-level tasks, such as triangles, cliques, and cycles.  The model allows the neural agents to retrieve information from the node they are occupying, their neighboring nodes (when they visit those nodes), and the co-located devices. 
They showed that this agent-based model was able to detect cliques and cycles, which was shown to be impossible in the widely-studied GNN architectures based on the message-passing model   (i.e., where devices are static and communication is through passing messages). 

Additionally, a recent study  \cite{AgentTriangleCounting} has shown that the  fundamental graph-level task of  {\em triangle detection} 
can be solved in the agent-based model by a deterministic algorithm in $O(\Delta \log n)$ rounds with $O(\Delta \log n)$ bits at each device.
In contrast, it is known that in the {\sf CONGEST} message-passing model it takes $O(n^{1/3}\polylog(n))$ rounds to solve triangle detection by a randomized algorithm \cite{ChangS19}, which is almost tight since there is the $\Omega(n^{1/3}/\log n)$ lower bound \cite{IzumiG17,Pandurangan0S18}, and hence 
the agent-based model provides a clear advantage when $\Delta<n^{1/3}\polylog(n)$ despite restriction on communication through device relocation. 

\subsection{Computing Model} 
We model the network as a connected, undirected graph $G=(V,E)$ with $|V|=n$ nodes and $|E|=m$ edges.  Each node $v_i \in V$ has $\delta_i$ ports corresponding to each edge incident to it labeled in $[1, \ldots, \delta_i]$. 
We assume that the set of $\cR=\{r_1,r_2,\ldots,r_n\}$ of $n$  agents are initially positioned on the nodes of $G$\footnote{Some graph problems may be solved with $k<n$ agents but not all problems such as MST, i.e., MST computed with $k<n$ agents will not be the MST of $G$ but of a sub-graph of $G$. Additionally, having $|\cR|=|V|=n$  agents makes the agent-based model equivalent to  $|V|=n$ devices (one device per node) in the message-passing model.}. The agents have unique IDs in the range $[1,n^{O(1)}]$.
The agents neither know the topology of $G$ nor the graph parameters (such as the network size $n$, maximum degree $\Delta$, diameter $D$, etc.). 
Initially, a node may have zero, one, or multiple agents positioned. An agent at a node can communicate with some (or all) agents present at that node, but not with the agents that are situated at some other nodes (a.k.a. the {\em local communication} model). 

An agent can move from node $v$ to node $u$ along the edge $e_{vu}$. 
Following the message-passing literature, e.g., \cite{GarayKP93}, we assume that an agent can traverse an edge in a round, irrespective of its (edge) weight even when $G$ is weighted. 
An agent that moves from $v$ along the port $p_{vu}$ is aware of the port $p_{uv}$ when it arrives at $u$.
Additionally, at any node $v$, it is aware of the weight $w(e)$ (if $G$ is weighted) of the edge $e_{vu}$ that connects $v$ to its neighbor $u$. We assume that there is no correlation between two port numbers of an edge. 
Any number of agents are allowed to move along an edge at any time, that is the agent-based model does not put restrictions on  
how many agents can traverse an edge at a time. 

The agents use the synchronous setting as in the standard $\mathcal{CONGEST}$ model: In each round, each agent $r_i$ positioned at a node $v_i$ can perform some local computation based on the information available in its storage as well as the (unique) port labels at the positioned node and decide to either stay at that node or exit it through a port to reach a neighbor node. Before exiting, the agent might write information on the storage of another agent which is staying at that node.   An agent exiting a node always ends up reaching another node (i.e., an agent is never positioned on an edge at the end of a round).  
We assume that the agents wake up simultaneously at the beginning of the execution. The time complexity is the number of rounds of operations until a solution. The storage (memory) complexity is the number of bits of information stored at each agent throughout the execution.  

Notice that at any round, the agent positions on $G$ may satisfy the following: 
\begin{itemize}
\item {\em dispersed} -- $n$ agents are on $n$ nodes,  
\item {\em rooted} -- $n$ agents are at a single node, or 
\item {\em general} -- neither {\em dispersed} nor {\em rooted}. 
\end{itemize}
If the agents are in dispersed (respectively, rooted, general) configuration initially, then we say the agents satisfy a dispersed (respectively, rooted, general) initial configuration.  We say an agent is {\em singleton} if it is alone at a node, otherwise {\em non-singleton} (or multiplicity).

\subsection{Contributions}
\label{sec:contribution}

\begin{table*}[ht]
\centering
{\footnotesize
\begin{tabular}{|c|ccc|c c|}
\toprule
\multirow{2}{*}{problem} &  \multicolumn{3}{c|}{previous result} &  \multicolumn{2}{c|}{our result (no parameter known)}\\ \cline{2-4} \cline{5-6}
  & time & memory/agent & known & time & memory/agent    \\ 
\toprule
leader & $-$ & $-$& $-$ & $O(m)$ & $O(n\log n)$ (Thm.~\ref{theorem:leader})  \\
 &  & &  &  & $O(\log^2 n)$ (Thm.~\ref{theorem:leader}, $\mathcal{D}$)  \\
\hline
MST & $-$ & $-$ & $-$ & $O(m+$ & $O(n\log n)$ (Thm.~\ref{theorem:result})  \\
 &  & &  & $n\log n)$ & $O(\log n\min\{\Delta,\log n\})$ (Thm.~\ref{theorem:result}, $\mathcal{D}$)  \\
\hline
gathering & $O(n^3)$ &  $O(M+ m \log n)$ & $n$ \cite{Molla-2021IPDPS} & $O(m)$ & $O(n\log n)$  (Thm.~\ref{theorem:Gathering}) \\ 
 &  & &  &  & $O(\log^2 n)$ (Thm.~\ref{theorem:Gathering}, $\mathcal{D}$)  \\
\hline
MIS & $O(n\Delta\log n)$ & $O(\log n)$ & $n,\Delta$ \cite{PattanayakBCM24} & $O(n\Delta)$&$ O(n\log n)$ (Thm.~\ref{theorem:MIS}) \\ 
 &  & &  &  & $O(\log^2 n)$ (Thm.~\ref{theorem:MIS}, $\mathcal{D}$)  \\
\hline
MDS & $O(\gamma \Delta \log n + $&$ O(\log n) $&$ n, \Delta, $  & $O(m)$&$ O(n\log n)$ (Thm.~\ref{theorem:MDS})\\
&  $n\gamma+m)$& & $m,\gamma$ \cite{ChandMS23} & & $O(\log^2 n)$ (Thm.~\ref{theorem:MDS}, $\mathcal{D}$)\\
\bottomrule
\end{tabular}
\caption{Summary of previous and our results in the agent-based model. $M$ is the memory required for the Universal Exploration Sequence (UXS) \cite{Ta-ShmaZ14} and $\gamma$ is the number of clusters of agents in the initial configuration. 
Previous results have parameter assumptions as outlined above. Our results do not have such assumptions. `$-$' means no previous result for the corresponding problem.  `$\mathcal{D}$' denotes the dispersed initial configuration. 
\label{table:results}
}}
\end{table*}

The first natural question to ask is: {\em Can algorithms for a problem in the message-passing model be simulated to solve that problem in the agent-based model?} This will allow bringing the vast literature on solving distributed graph problems in the message-passing model to the agent-based model. 
We show that this is indeed possible, however under certain assumptions as  
outlined in the following general theorem. Therefore, the goal of this paper is to lift the assumptions by devising algorithms directly in the context of the agent-based model.   

\begin{theorem}[{\bf Simulation}]
\label{theorem:coversion}
Any deterministic algorithm  $\mathcal{A}$ 
that takes $O(T)$ rounds in the message-passing model can be converted to a deterministic algorithm $\mathcal{A'}$ that takes 
\begin{itemize}
    \item $O(\Delta T \log n)$ rounds, if dispersed initial configuration
    \item $O(\Delta T \log n + n \log^2 n)$ rounds, rooted or general initial configuration 
\end{itemize}
in the agent-based model, provided that graph parameters $n$ and $\Delta$ are known to agents a priori and memory at each agent is as much as the node memory needed for algorithm   $\mathcal{A}$ in the message-passing model. 
\end{theorem}

If we consider the $T=O(D)$-round time-optimal deterministic algorithm for leader election in the message-passing model \cite{Peleg90L}, a leader can be elected in the agent-based model in time  $O(\Delta D \log n + n \log^2 n)$ rounds, where $D$ is the diameter of $G$. An MST of $G$ can be constructed in $O(\max\{\Delta \sqrt{n} \log n \log^*n, \Delta D \log n,n\log^2n\})$ rounds in the agent-based model simulating the famously known as GKP algorithm \cite{GarayKP93,KuttenP98} that takes    
$T=O(\sqrt{n}\log ^{*}n+D)$ rounds in the message-passing model. 

Given Theorem \ref{theorem:coversion}, the immediate overarching question to ask is:  {\em Can the assumptions of known graph parameters (such as $n$ and $\Delta$) in Theorem \ref{theorem:coversion} be lifted and 
solutions obtained with limited memory per agent?} 
In this paper, we show for the very first time that this is indeed possible in the agent-based model. We do so by developing a deterministic algorithm for leader election in which, without agents knowing any graph parameter a priori, one agent is elected as a leader. Specifically, we prove the following theorem. 

\begin{theorem}[{\bf Leader election}]
\label{theorem:leader}
There is a deterministic algorithm in the agent-based model that elects one agent at a graph node as a leader in $O(m)$ rounds with
\begin{itemize}
    \item $O(\log^2 n)$ bits per agent, if dispersed initial configuration,
    \item $O(n \log n)$ bits per agent, if rooted or general initial configuration,
    \end{itemize}
without agents knowing any graph parameter (neither $n$ nor $\Delta$) a priori. 
\end{theorem}

As a byproduct, Theorem \ref{theorem:leader} guarantees that $n$ agents reach a dispersed configuration if they were not initially. This byproduct was considered as an independent problem of dispersion in the agent-based model (where the goal was to reach a dispersed configuration starting from any initial configuration) and studied in a sequence of papers (e.g., \cite{kshemkalyani2018,KshemkalyaniMS19,KshemkalyaniICDCS20,KshemkalyaniMS22,KshemkalyaniS21-OPODIS}).   The leader election guarantee in Theorem \ref{theorem:leader} is {\em implicit} meaning that the leader agent knows it is a leader and all others know they are not leaders but non-leaders do not know which agent is the leader.  
The leader election result as well as the byproduct result are interesting and important. We show why so by first establishing the following theorem for MST. 



\begin{theorem}[{\bf MST}]
\label{theorem:result}
There is a deterministic algorithm for constructing MST in the agent-based model that takes $O(m+n\log n)$ rounds and 
\begin{itemize}
    \item $O(\log n \cdot \min\{\Delta,\log n\})$ bits per agent, if dispersed initial configuration,
    \item $O(n \log n)$ bits per agent, if rooted or general initial configuration,
    \end{itemize}
without agents knowing any graph parameter a priori.
\end{theorem}

Compared to the time bound for MST obtained through simulation (Theorem \ref{theorem:coversion}) with assumptions on graph parameters, our MST result (Theorem \ref{theorem:result}) is better for any $G$ where $m<\max\{\Delta \sqrt{n} \log^*n,\Delta D, n\log n\}$, without agents knowing any graph parameter a priori.  

Finally, we show the importance of the leader election result by 
improving the time and/or memory complexities of the existing results on gathering \cite{Molla-2021IPDPS}, maximal independent set (MIS) \cite{PattanayakBCM24}, and minimal dominating sets (MDS) \cite{ChandMS23} in the agent-based model, lifting the assumptions of known graph parameters.   These gathering, MIS, and MDS results are discussed in detail in Section \ref{section:applications}. 
Table \ref{table:results} summarizes the results in the agent-based model. 

\subsection{Challenges}
The message-passing model allows the nodes (processors) to send/receive messages to/from their neighbors, i.e.,  in a single round, a node can send a message to all its neighbors and receive messages from all its neighbors. 
In contrast, in the agent-based model, the messages from an agent, if any, that are to be sent to the other agents in the neighboring nodes have to be delivered by the agent visiting those neighbors. Furthermore, it might be the case that when the agent reaches that node, the agent at that node may have already moved to another node. Therefore, any algorithm in the agent-based model needs to guarantee message delivery by synchronizing sender and receiver agents to be co-located at a node. 

Additionally, the graph-level tasks (such as MST) demand each node of $G$ to have an agent positioned on it to be able to provide a solution, i.e., if agents are not in a dispersed configuration, 
then MST constructed may not the MST of whole $G$ but its sub-graph.  Additionally, the MST computed may be the MST forest of graph components formed by agent positions.  Notice that the initial configuration of $n$ agents in a $n$-node graph $G$ may not be dispersed. 

Suppose initially the agent configuration is dispersed. 
Surprisingly, even in this initial configuration, the agent positioned at a node does not know this configuration.  Therefore, irrespective of whether the nodes have zero, single, or multiple agents initially, it seems highly advantageous to reach a dispersed configuration. 

Suppose the agents are in a dispersed configuration  
and the goal is to construct MST. The question is which agent starts MST construction and when. 
The leader election problem handles this symmetry breaking issue, since if a leader can be elected, then the authority can be given to the leader agent to initiate MST construction.   
The remaining agents do not participate in MST construction until the leader grants them authority to do so. Although having a leader seems to make MST construction easier and possibly other problems too, electing a leader itself 
turned out to be a difficult task.

\subsection{Outline of Techniques}
We develop a technique to elect a leader in two stages. In the first stage, the agents compete to become `local leader'. In the second stage, the local leaders compete to become unique `global leader'. The agent becoming a local leader can immediately run the second stage.

Once becoming a local leader, each local leader can run the same {\em global election} procedure in the second stage to become a global leader. However, in the first stage, an agent becomes a `local leader' through one of the two procedures. 
If an agent is initially alone, it will run a {\em singleton election} procedure to become a local leader. %
If an agent is not initially alone, it runs a {\em multiplicity election} procedure to become a local leader.   
The singleton election procedure run by an agent $r_v$ at a node $v$ visits the neighbors of $v$ (possibly repeatedly).
The multiplicity election procedure run by an agent $r_v$ at a node $v$ may possibly visit many nodes of $G$, possibly non-neighbors of $v$, placing an agent on each empty node visited. 
The singleton election procedure elects an agent $r_v$ at node $v$ as a local leader if and only if all the neighbors of $v$ have a singleton agent positioned initially. That is, if the procedure finds at least a neighbor of $v$ is empty or an agent from the multiplicity election procedure positioned, it will not elect $r_v$ as a leader. 
The multiplicity election procedure tries to elect an agent which becomes singleton at the end as a local leader.
The proposed technique guarantees that starting from any initial configuration, at least one agent becomes a local leader.


For the dispersed initial configuration, only the singleton election procedure runs. For the rooted initial configuration, only the multiplicity election procedure runs. 
For the general initial configuration,  singleton election and/or multiplicity election procedures run concurrently.  
Once all multiplicity election procedures starting from different nodes finish, it is guaranteed that agents are in a dispersed configuration, a byproduct result. 

As soon as some agent becomes a local leader, it runs the global election procedure to become a global leader. The global election procedure is to check whether the local leader agent will be able to traverse all the edges of $G$. 
If the local leader agent is indeed able to traverse all the edges of $G$, it returns to the node where it became local leader (which we call the {\em home node} of that local leader) 
and declares itself as a global leader. We prove that after a local leader agent elects itself as a global leader, there is no other local leader agent that can elect itself as a global leader (i.e., the global leader is unique). 

Since there might be multiple local leaders elected in the first stage, there might be multiple global election procedures running concurrently. Each global election procedure $p$ has tuple $(roundNo_p, ID_p)$ such that if it meets another global election procedure $(roundNo_q,ID_q)$ then $p$ continues and $q$ stops if $p$'s tuple is lexicographically greater than $q$'s, otherwise $q$ continues and $p$ stops. Here $roundNo_*$ is a round number at which the procedure started and $ID_*$ is the ID of the agent that runs this procedure.

Consider the home nodes 
of the local leaders before they run the global election procedure. During local leaders run the global election procedure, those home nodes become empty since the agents are traversing $G$. When some other global election or multiplicity election procedure encounters an empty node, it needs to confirm whether the empty node is, in fact, empty or a home node of some local leader that is not currently at the home node. This is done by asking the local leaders to keep the home node information at an agent positioned at a neighbor node. 
The global election procedure visits the neighbors (of an empty node) to see whether such home node information exists at a neighbor. We prove that if an empty node is indeed a home node then there exists a neighbor holding that information.  Additionally, we prove that 
the local leader can return to its home node anytime it desires.

After a leader is elected, as an application, we use it to solve other fundamental problems. One is MST construction which was not considered in the agent-based model before. The rest are gathering, MIS, and MDS problems which were considered in the agent-based model before but solved assuming that the agents know one or more graph parameters a priori. We lift those assumptions and additionally provide improved time/memory bounds.  

For the MST construction, the leader plays a crucial role in synchronizing the agents. The leader ranks the agents and starts constructing an MST. It keeps its rank the highest. The leader, once its job is done, informs that second ranked agent to continue constructing MST. The second informs the third, and so on, until $(n-1)$-ranked agents pass the token to the $n$-th ranked. The $n$-th ranked agent passes the token back to the leader and one phase of MST construction finishes. 
It is guaranteed that at the end of this phase, there will be at least $n/2$ edges of the MST identified. 
Therefore, repeating this process for  $O(\log n)$ phases, we have all $n-1$ edges of MST correctly identified, giving an MST of $G$. 
\subsection{Related Work}

The leader election problem was first stated by Le Lann \cite{Lann77} in the context of token ring networks, and since then it has been central to the theory of distributed computing.  Gallager, Humblet, and Spira \cite{Gallager83} provided a deterministic algorithm for any $n$-node graph $G$ with time complexity $O(n\log n)$ rounds and message complexity $O(m\log n)$. Awerbuch \cite{Awerbuch87} provided a deterministic algorithm with time complexity $O(n)$ and message complexity $O(m+n\log n)$. Peleg \cite{Peleg90L} provided a deterministic algorithm with optimal time complexity $O(D)$ and message complexity $O(mD)$. Recently, an algorithm is given in \cite{KPP0T15} with message complexity $O(m)$ but no bound on time complexity, and another algorithm with $O(D\log n)$ time complexity and $O(m\log n)$ message complexity. Additionally, it was shown in \cite{KPP0T15} the message complexity lower bound of $\Omega(m)$ and time complexity lower bound of $\Omega(D)$ for deterministic leader election in graphs.  Leader election was not studied in the agent-based model before.  

For MST,  the algorithm in Gallager, Humblet, and Spira  \cite{Gallager83} 
is the first deterministic algorithm in the message-passing model with  
time complexity $O(n\log n)$ and message complexity $O(m\log n)$. Time was improved to 
$O(n)$ in \cite{Awerbuch87}  and to 
$O(\sqrt{n}\log ^{*}n+D)$ in \cite{GarayKP93,KuttenP98}. Furthermore, a time lower bound of 
$\Omega (\sqrt {n}/{\log n}+D)$ was given in \cite{PelegR00}. MST was not studied in the agent-based model before. 

For MIS in the message-passing model, the best-known deterministic distributed algorithms have time complexity $O(2^{\sqrt{\log n}})$ \cite{AGLP1989,PS1996}. For MDS, Deurer {\it et al.} \cite{DKM2019} gave two algorithms with an optimal approximation ratio of $(1+\epsilon)(1+\log(\Delta+1))$ running respectively in $O(2^{O(\sqrt{\log (n)\log(\log (n)})})$ and $O(\Delta \polylog(\Delta)+\polylog(\Delta)\log^\star (n))$ rounds for $\epsilon>\frac{1}{\polylog(\Delta)}$. MIS and MDS were solved in the agent-based model in \cite{ChandMS23,PattanayakBCM24} with time and memory complexities reported in Table \ref{table:results} assuming $n,\Delta$ (additionally $m,\gamma$ for MDS) are known to agents a priori. We improve these results w.r.t. time/memory complexities as well as lift the parameter assumptions.  

Gathering is a very old problem and has been studied extensively in the agent-based model. The recent results are  \cite{Molla-2021IPDPS,Ta-ShmaZ14} (detailed literature in \cite{Molla-2021IPDPS}). \cite{Ta-ShmaZ14} provided a $\tilde{O}(n^5\log \beta)$ time solution to gather $k\leq n$ agents in arbitrary graphs, where $\tilde{O}$ hides polylog factors and $\beta$ is the smallest label among agents. 
Molla {\it et al.} \cite{Molla-2021IPDPS} provided improved time bounds for large values of $k$ assuming $n$ is known but not $k$: (i) $O(n^3)$ rounds, if $k \geq \floor*{\frac{n}{2}} +1$ (ii) $\Tilde{O}(n^4)$ rounds, if $ \floor*{\frac{n}{2}} +1 \leq k < \floor*{\frac{n}{3}} +1$, and (iii) $\Tilde{O}(n^5)$ rounds, if $ \floor*{\frac{n}{3}} +1 > k$. Each agent requires $O(M + m \log n)$ bits of memory, where $M$ is the memory required to implement the universal traversal sequence (UXS) \cite{Ta-ShmaZ14}. 
We lift the assumption of known $n$ and improve time bound for $k=n$ case from $O(n^3)$ to $O(m)$ and additionally the memory bound (see Table \ref{table:results}).

Finally, there is a model, called {\em whiteboard} \cite{Sudo2010}, 
related to the agent-based model. The whiteboard model 
considers (limited) storage at the network nodes in addition to (limited) storage per device. Other aspects remain the same as in the agent-based model, for example, devices have identifiers, computation ability, and communication through relocation. 
It is immediate that any solution designed in the agent-based model works in the whiteboard model without any change but the opposite may not be true.

\subsection{Roadmap}
We establish our simulation result (Theorem \ref{theorem:coversion}) in Section \ref{sec: MPM_to_RM}. We then discuss our leader election algorithm in  
Section \ref{section:leader} proving Theorem \ref{theorem:leader}.  Using the elected leader,  we present our MST construction algorithm in Section \ref{sec: MST construction} proving Theorem \ref{theorem:result}. Hereafter, we provide improved results to the gathering, MIS, and MDS problems in Section \ref{section:applications}.   
Finally, we conclude in Section \ref{section:conclusion} with a discussion on memory requirement and on future work. 

\section{Simulating Message-Passing Algorithm in the Agent-based Model}\label{sec: MPM_to_RM}

In this section, we prove Theorem \ref{theorem:coversion} which establishes a time bound of a solution to a problem in the agent-based model simulating a deterministic solution to that problem from the message-passing model. Before we do that, we prove the following crucial lemma. 

\begin{lemma}\label{lem: meeting_with_ngbrs}
Given a dispersed configuration of $n$ agents on a $n$-node graph $G$,
    an agent at a node takes at most $O(\Delta \log n)$ rounds to meet all its neighbors,  provided that $n$ and $\Delta$ are known to agents a priori. 
\end{lemma}
\begin{proof}
    Since $n$ is known and the agent IDs are in  $[1,n^{O(1)}]$, each agent ID can be encoded by $c\cdot \log n=O(\log n)$ bits, for some constant $c$. 
    If the length of agent ID $x<c\cdot \log n$ bits, then all the $c\cdot \log n-x$ bits preceding the most significant bit (MSB) are filled with bit 0.  
    Each agent $r_v$ at node $v$ then runs a neighbor probe procedure, knowing $\Delta$, as follows. 
    Start from MSB and end at the padded bit one by one at the interval of $2\Delta$ rounds. Agent $r_v$ stays at $v$ for $2\Delta$ rounds when the bit accessed is $0$, otherwise (i.e., bit accessed is 1) $r_v$ visits all the neighbors of $v$ one by one. We know that $r_v$ can visit all its neighbors in $2\Delta$ rounds since visiting a neighbor needs 2 rounds. 
    Recall that each agent has a unique ID, therefore, the IDs of two agents $r_v,r_u$  possess at least one-bit difference in their IDs. This bit difference ensures two neighbors meet each other when their ID bits differ. 
    Since IDs of all the agents are of length $O(\log n)$ bits, the IDs of each of the neighbors must differ by at least a bit when $r_v$ visits all its neighbors for $O(\log n)$ time.  
    Therefore, each agent meets all of its neighbors in $O(\Delta \log n)$ rounds.
\end{proof}


\noindent{\bf Proof of Theorem \ref{theorem:coversion}:}

\begin{proof}
To simulate an algorithm in the message-passing model to the agent-based model, the agents need to be in a dispersed configuration, if they were not dispersed initially. The known time complexity of achieving dispersion configuration, starting from any initial general configuration, is $O(n\log^2n)$ rounds due to Sudo {\it et al.} \cite{sudo23}. 
The algorithm of Sudo {\it et al.} \cite{sudo23} is non-terminating meaning that each agent may not know when the dispersion configuration is achieved since $n$ is not known. Additionally, if $\Delta$ is not known, an agent may not be able to be co-located with its neighboring agent to deliver a message. 
Therefore, known $n$ helps to transition from the dispersion procedure to the algorithm simulation procedure. Knowing $\Delta$ is needed to establish  Lemma \ref{lem: meeting_with_ngbrs}. 


Consider memory at each agent the same as it would have been used at a node in the message-passing model, so that, for the received message at the node or agent, both models can perform the required computation. 
It is sufficient to prove that, after $O(\Delta \log n)$ rounds, the agents in the agent-based model have the same information as in the nodes in the message-passing model after executing any deterministic algorithm $\mathcal{A}$ for a round. 
%
The proof is as follows.
In any single round of the message-passing model, a node passes the message/information to its all (or fewer) neighbors. To perform the same operation in the agent-based model, it is essential that an agent meets all its neighbors and passes the message based on the algorithm. Since each agent meets all its neighbors in $O(\Delta \log n)$ rounds as shown in Lemma~\ref{lem: meeting_with_ngbrs}, a round in the message-passing model can be simulated in $O(\Delta \log n)$ rounds in the agent-based model. Thus, the total rounds required to simulate a deterministic algorithm $\mathcal{A}$ that runs for $O(T)$ rounds in the message-passing model is $O(\Delta T \log n)$ rounds in the agent-based model. 
Combining the time bounds to achieve a dispersion configuration (if the configuration initially is not dispersed) and the algorithm simulation, we have the claimed bound of $O(\Delta T \log n+n\log^2n)$ rounds for rooted and general initial configurations and simply $O(\Delta T \log n)$ rounds for dispersed initial configurations.
\end{proof}





\begin{algorithm}[bt!]
{
\footnotesize
\SetKwInput{KwInput}{Input}
\SetKwInput{KwStates}{States}
\SetKwInput{KwEnsure}{Ensure}
\KwInput{A set $\cR$ of $n$ agents with unique IDs positioned initially arbitrarily on the nodes of an $n$-node, $m$-edge anonymous graph $G$.}

\KwEnsure{An agent in $\cR$ is elected as a leader with status $leader$.}          

\KwStates{Initially, each agent $r_u$ positioned at node $u$ has $r_u.status\leftarrow candidate$, $r_u.init\_alone \leftarrow true$ if alone at $u$,  $r_u.init\_alone \leftarrow false$ otherwise, and $r_u.all\_edges\_visited\leftarrow false$. The $init\_alone$ variable is never updated for $r_u$ throughout the algorithm but the $status$ variable can take values $\in\{non\_candidate, local\_leader, leader\}$ and the $all\_edges\_visited$ variable can take value $true$.  
} 












\If{$r_u.status=candidate$}
{   

    
    \If{$r_u.init\_alone=True$}
    { 
        $Singleton\_Election(r_u)$
    }
    \If{$r_u.init\_alone = false$}
    {
    \If{$r_u$ is the minimum ID agent at $u$}
    {
        $Multiplicity\_Election(r_u)$
    }
}}
\If{$r_u.status=local\_leader$}
{
\If{$r_u$ became local leader through $Multiplicity\_Election(r_u)$}
{Ask the parent node, say $w$,  in the DFS tree built while running $Multiplicity\_Election(r_u)$ to keep the information that node $u$ is the {\em home} node of the local leader $r_u$. }
\If{$r_u$ became local leader through $Singleton\_Election(r_u)$}
{
Inform all neighbors that $r_u$ is a local leader.

                Ask the port-1 neighbor, say $w$,   to keep the information that node $u$ is the {\em home} node of the local leader $r_u$. 
}
$Global\_Election(r_u)$

\If{$r_u$ is at the home (or root) node from where Global\_Election($r_u$) started with $r_u.all\_edges\_visited=true$ }
{
$r_u.status\leftarrow leader$
}
} 
}
\caption{Leader election for agent $r_u$}
\label{algorithm:leader_election}
\end{algorithm}

\begin{algorithm}[bt!]
{


\footnotesize
$\delta_u\leftarrow$ degree of node $u$.

$N(r_u) \leftarrow$ neighbors of agent $r_u$.

$r_u$ visits neighbors in $N(r_u)$ (in order of increasing port numbers) one by one starting from and ending at $u$.
 
\While{$r_u.status==candidate$}
{

    \If {$\exists$ neighbor $v, \delta_u>\delta_v$ or at least a neighboring agent belongs to Multiplicity\_Election() or has status local\_leader or ( $\exists$ neighbor $v$, $\delta_u=\delta_v$ such that $r_v.ID>r_u.ID$) 
    }
    {
        $r_u.status\leftarrow non\_candidate$
    }
        
    \ElseIf{$\exists$ neighbor $v$, $\delta_u=\delta_v$ and $v$ is empty 
    \label{line: start_same_degree_leader}}
    {
        $Neighbor\_Exploration\_with\_Padding(r_u)$  
    }

    \ElseIf{$\forall$ neighbor $v$, $\delta_v>\delta_u$ but $\exists$ (at least) a neighbor $v'$ which is empty} 
    {
        $r_u$ visits the empty neighbors in the interval of  $2\delta_{v'}$ rounds starting from and ending at $u$.
        
        \If{an agent $r_{v'}$ is found at $v'$ and $r_v'$ belongs to Multiplicity\_Election() or has status local\_leader} 
        {
            $r_u.status\leftarrow non\_candidate$
        }
    }
    \If {$\forall$ neighbor $v, \delta_u<\delta_v$ and all neighboring agents were initially singletons and no neighbor has status local\_leader and ( $\forall$ neighbor $v$, if $\delta_u=\delta_v$ then $r_v.ID<r_u.ID$) 
    }
    {
        $r_u.status\leftarrow local\_leader$
    }                    
}\label{line: end_same_degree_leader}
        
} 

        \caption{$Singleton\_Election(r_u)$}
\label{algorithm:local_leader_election}
\end{algorithm}

\begin{algorithm}[bt!]
{
\footnotesize
$b\leftarrow$ number of bits in the ID of $r_u$

$b+2b^2 \leftarrow$ number of bits in the ID of $r_u$ after padding a sequence of `10' bits $b^2$ times to the LSB in the original $b$-bit ID.

Starting from MSB and ending on LSB, if the bit is '1' visit the $N(r_u)$ one by one which finishes in $2\delta_u$ rounds. If bit is '0' stay at $u$ for $2\delta_u$ rounds. 

$r_u$ explores $N(r_u)$ based on padding for $2\delta_u (b+2b^2)$ rounds

    \If{$r_u$ meets an agent $r_v$ running Algorithm~\ref{algorithm:dispersion} ($Multiplicity\_Election$) or Algorithm~\ref{algorithm:leader-election} ($Global\_Election$) or ( $\exists$ neighbor $v$, $\delta_u=\delta_v$ such that $r_v.ID>r_u.ID$)}  
    {
        $r_u.status\leftarrow non\_candidate$.
    }

\If { $\exists$ neighbor $v$, $\delta_u=\delta_v$ and $v$ is empty} 
{   
   $r_u.status\leftarrow non\_candidate$.
}
        }
        \caption{$Neighbor\_Exploration\_with\_Padding()$}
 \label{algorithm:padding}
\end{algorithm}

\begin{algorithm}[bt!]
{
\footnotesize
        Run Depth First Search (DFS) traversal in forward and backtrack phases.  The DFS has ID $r_u$ and can be denoted as $DFS(r_u)$. The agents initially co-located with $r_u$ move with $r_u$ as a group. In each empty node, say $v$, visited, $DFS(r_u)$ waits for a round. If $v$ is still empty after the (waited) round,  $DFS(r_u)$ runs the $Confirm\_Empty()$ procedure (Algorithm \ref{algorithm:confirm-empty}) to verify $v$ is in fact empty. If the $Confirm\_Empty()$ procedure verifies $v$ empty, $DFS(r_u)$ asks the largest ID agent, say $r_v$, in its group to stay at $v$ setting $r_v.status\leftarrow non\_candidate$. $DFS(r_u)$ runs until $r_u$ reaches to a confirmed empty node $w$ alone where it can change its status from candidate to local leader setting $r_u.status\leftarrow local\_leader$ (the node $w$ becomes the {\em home} node for local leader $r_u$). If the head of $DFS(r_x)$ meets the head of $DFS(r_y)$ at a node $w$, then the highest ID agent in the group belonging to $DFS(r_x)$ stays at $w$ (and becomes $non\_candidate$) if $r_x>r_y$, otherwise the highest ID agent in the group belonging to $DFS(r_y)$ stays at $w$ (and becomes $non\_candidate$). Both $DFS(r_x)$ and $DFS(r_y)$ continue their traversal until their respective heads become singleton at empty nodes and the heads are elected as local leaders.

        {\it {\bf Remark:} There might be a situation in which when $Multiplicity\_Election()$ finishes for an agent $r_w$ at a node $w$, the parent node $w'$ of $w$ in the DFS tree built by $r_w$ may be empty (this is because the parent node happened to be a home node of a local leader $r_{w'}$ that is currently running $Global\_Election$). In this case, $r_w$ returns to $w'$ from $w$ and waits there until $r_{w'}$ returns to $w'$. Once $r_w$ meets $r_{w'}$ (at $w'$), $r_{w'}$ becomes ``non-candidate'' and stays at $w'$. $r_w$ goes to node $w$ from $w'$ and becomes a local leader if it finds no waiting agent at $w$. If there is a waiting agent at $w$, $r_w$ becomes a ''non-candidate'' and the waiting agent leaves $w$. This process continues until a waiting agent can become a local leader with the parent node in its DFS tree non-empty.   
        } 
        }
        \caption{$Multiplicity\_Election(r_u)$}
         \label{algorithm:dispersion}
\end{algorithm}

\begin{algorithm}[bt!]
{
\footnotesize
        This procedure is to verify whether the empty node, say $x$, encountered by procedure $Multiplicity\_Election()$ (Algorithm \ref{algorithm:dispersion}) or $Global\_Election()$ (Algorithm \ref{algorithm:leader-election}) is in fact empty, is the {\em home} node of a local leader, or a possible home node of an agent waiting to become a local leader (see Line 2 of Algorithm \ref{algorithm:dispersion}).  This is done as follows. The agent running this procedure at an empty node $x$ visits the neighbors $N(x)$ of $x$ and collects information on whether one of the neighbors has the information that $x$ is the (home) node of a (possible) local leader. If no neighbor has the information that $x$ is a (possible) home node, it is verified that $x$ is empty, otherwise, it is a (possible) home node.  
        
        {\it {\bf Remark:} There may be the case that procedure  ($Multiplicity\_Election()$ or $Global\_Election()$) from a different agent may reach $x$ while one agent is running this procedure at $x$. These procedures may simply wait at $x$  for the ongoing $Confirm\_Empty()$ procedure to finish.  
        } 
        }
        \caption{$Confirm\_Empty()$}
 \label{algorithm:confirm-empty}
\end{algorithm}

\begin{algorithm}[bt!]

 {
 \footnotesize
        Run DFS traversal in forward and backtrack phases as in $Multiplicity\_Election(r_u)$ (Algorithm \ref{algorithm:dispersion}).  The DFS has ID as a tuple $(roundNo_u,r_u)$ and it can be denoted as $DFS(roundNo_u,r_u)$, where $roundNo_u$ denotes the round at which this DFS has started. $DFS(roundNo_u,r_u)$ starts at the home node of $r_u$ where it becomes a local leader and ends at the home node (the home node is also the root node of $DFS(roundNo_u,r_u)$). 
        The goal of $DFS(roundNo_u,r_u)$ is to see whether it can visit each and every edge of $G$. $DFS(roundNo_u,r_u)$ keeps a boolean variable $alledgevisited$ initially set to $False$. As soon as no edge is left to be visited, $allegdgevisited$ becomes $True$ and if $r_u$ is not on the home (or root) node from where $DFS(roundNo_u,r_u)$ started, it comes back to that node following the parent pointers.  
        
        While running $DFS(roundNo_u,r_u)$, if $r_u$ reaches a node in which it meets an agent running $Multiplicity\_Election()$, $DFS(roundNo_u,r_u)$ continues.  However, if $DFS(roundNo_u,r_u)$ meets another $DFS(roundNo_v,r_v)$ at a node $w$ (not necessarily the head of $DFS(roundNo_v,r_v)$), $DFS(roundNo_u,r_u)$ continues if and only if  either (i) $roundNo_u>roundNo_v$ or (ii) $roundNo_u=roundNo_v$ but $r_u>r_v$, otherwise $DFS(roundNo_u,r_u)$ stops and $DFS(roundNo_v,r_v)$ continues. 
        If $DFS(roundNo_u,r_u)$ stops, then agent $r_u$ follows its parent pointers to reach its home (root) node. 

        While running $DFS(roundNo_u,r_u)$, If $r_u$ reaches an empty node, it waits at that node for a round, and if the node is still empty after the (waited round), runs procedure $Confirm\_Empty()$  as in  $Multiplicity\_Election(r_u)$ (Algorithm \ref{algorithm:dispersion}) to verify whether the empty node is a home node (of a local leader). If so, $DFS(roundNo_u,r_u)$ continues, otherwise $DFS(roundNo_u,r_u)$ stops and $r_u$ returns to its home node.  
}
   \caption{$Global\_Election(r_u)$}
    \label{algorithm:leader-election}       
\end{algorithm}

\section{Leader Election} 
\label{section:leader}
In this section, we present our deterministic leader election algorithm which, starting from any initial configuration (dispersed, rooted, or general) of $n$ agents on an $n$-node graph $G$, ensures that one agent is elected as a global leader. 
Additionally, as a byproduct, if the agents were initially in rooted or general configurations, they are in a dispersed configuration when the leader election finishes. 
%
We start with a high-level overview of the algorithm, then specific details, and finally the correctness and complexity proofs of the algorithm guarantees. 

\subsection{High-Level Overview of the Algorithm} Initially, a graph node may have zero, one, or multiple agents. All these agents are ``candidates'' to become leader. A candidate needs to first become a ``local leader'' before becoming a ``global leader''. Each candidate that cannot become a ``local leader'' (also each ``local leader'' that cannot become a ``global leader'') will become a ``non\_candidate''.  Lines 2-6 of Algorithm \ref{algorithm:leader_election}  show what procedure an agent runs to become a local leader.  

As depicted in Lines 2-3 of Algorithm \ref{algorithm:leader_election}, if an agent is initially singleton at a node, then it runs Algorithm \ref{algorithm:local_leader_election} ($Singleton\_Election$) to become a local leader. As depicted in Lines 4-6 of Algorithm \ref{algorithm:leader_election}, if an agent is not initially non-singleton then it runs Algorithm \ref{algorithm:dispersion} ($Multiplicity\_Election$) to become a local leader. 
After an agent becomes a local leader, 
it runs Algorithm \ref{algorithm:leader-election} ($Global\_Election$) to become a global leader.  

An agent $r_u$ running Algorithm \ref{algorithm:local_leader_election} ($Singleton\_Election$) at a node $u$  will be successful in becoming a local leader if and only if all $u$'s neighbors have initially a single agent positioned on them and $u$ has the smallest degree compared to the neighboring nodes. 
Each initially singleton agent $r_u$ at node $u$ running $Singleton\_Election$ visits the neighbors of $u$ one by one which finishes in $2\delta_u$ rounds, where $\delta_u$ is the degree of $u$.  If a subsequent visit is needed, then it will take again $2\delta_u$ rounds. If not all neighbors have initially singleton agents positioned,  the agent gets to know it cannot become a local leader. 
It then stops the algorithm and becomes ``non\_candidate''.

An agent $r_u$ initially at node $u$ running Algorithm \ref{algorithm:dispersion} ($Multiplicity\_Election$) will be successful in becoming a local leader if and only if it has the smallest ID among the ones positioned with it initially at $u$. As soon as the smallest ID agent becomes a singleton at node $w$, it declares itself as a local leader if the parent node in its DFS tree built while running $Multiplicity\_Election$ is non-empty (see Line 2 - Remark in Algorithm \ref{algorithm:dispersion}). 
If such a parent is empty, it waits at that parent to decide later whether to become a local leader or a non-candidate. 
Algorithm \ref{algorithm:dispersion} ($Multiplicity\_Election$) for an initially non-singleton agent $r_u$ with $\alpha$ co-located agents is a {\em Depth First Search} (DFS) traversal with the goal to visit $\alpha-1$ other empty nodes of $G$ on which $\alpha-1$ agents can stay and $r_u$ becomes a local leader. All other agents initially co-located with $r_u$ at node $u$ stay one by one on the empty nodes of $G$ visited by Algorithm \ref{algorithm:dispersion} 
and become ``non\_candidate''.

To make sure that $Multiplicity\_Election$ meets $Singleton\_Election$ (if it is running), $Multiplicity\_Election$ waits at a node for a round.  
$Singleton\_Election$ stops and the agent becomes non\_candidate when it knows about  $Multiplicity\_Election$.  

After becoming a local leader (irrespective of whether through $Singleton\_Election$ or $Multiplicity\_Election$), the local leader agent runs Algorithm \ref{algorithm:leader-election} ($Global\_Election$) to become a global leader. Algorithm \ref{algorithm:leader-election} ($Global\_Election$)  is a DFS traversal as in Algorithm \ref{algorithm:dispersion} ($Multiplicity\_Election$) with the goal to visit all the edges of $G$. 
To make it easier for other local leaders or $Multiplicity\_Election$ instance from another agent to not mistakenly put an agent on the home node of a local leader 
the neighbor nodes are asked to store the information about a home node. The agents running Algorithms \ref{algorithm:dispersion} and \ref{algorithm:leader-election}  check the neighbors to confirm whether the visited empty node is in fact a home node of a local leader (or a node of an agent that is waiting at a parent node to possibly become a local leader, see Line 2 of Algorithm  \ref{algorithm:dispersion}). The confirmation procedure is described in Algorithm \ref{algorithm:confirm-empty} ($Confirm\_Empty$). If an empty node is a home node (or possible home node of an agent waiting to possibly become a local leader), Algorithms \ref{algorithm:dispersion} and \ref{algorithm:leader-election} continue leaving that node empty as is. Otherwise, Algorithm \ref{algorithm:dispersion} puts an agent and continues, and Algorithm \ref{algorithm:leader-election} stops as it knows that $Multiplicity\_Election$ instance from at least one agent has not finished yet. 

There may be the case that while running Algorithm \ref{algorithm:leader-election}, $DFS(roundNo_{i},r_i)$ of local leader $r_i$ may {\em meet} $DFS(roundNo_j,r_j)$ of local leader $r_j$. In this case,  $DFS(roundNo_{i},r_i)$ continues if $roundNo_{i}>roundNo_{j}$ (if same round number, use agent IDs), otherwise $DFS(roundNo_{j},r_j)$. If  $DFS(roundNo_{j},r_j)$ stops, then $r_j$  becomes ``non\_candidate" and returns to its home node following parent pointers in $DFS(roundNo_{j},r_j)$. 

\subsection{Detailed Description of the Algorithm}
We discuss $Singleton\_Election$ (Algorithm \ref{algorithm:local_leader_election}), $Multiplicity\_Election$ (Algorithm \ref{algorithm:dispersion}),  and $Global\_Election$ (Algorithm \ref{algorithm:leader-election}) procedures, including how they synchronize when they meet the same or different procedure running concurrently by other agents (local leader or candidates) to decide on when to proceed and when to stop. 

\vspace{2mm}
\noindent{\bf Singleton\_Election (Algorithm \ref{algorithm:local_leader_election}).} This procedure is run by agents that were initially singleton on a node. They set status $candidate$ and run the $Singleton\_Election$ algorithm. The agent $r_u$ at node $u$ visits the $\delta_u$ neighbors of $u$ one by one starting from the minimum ID port to the maximum ID port. Agent $r_u$ finishes visiting all $\delta_u$ neighbors in $2\delta_u$ rounds. 
If $r_u$ finds there is a neighboring node $v$ such that $\delta_u>\delta_v$, $r_u$ becomes $non\_candidate$ (Lines 5 and 6 of Algorithm \ref{algorithm:local_leader_election}).
If $r_u$ finds there is at least a neighboring agent settled through $Multiplicity\_Election()$ or has status $local\_leader$, $r_u$ cannot become a local\_leader and hence sets its status $non\_candidate$ (Lines 5 and 6 of Algorithm \ref{algorithm:local_leader_election}). 
If all neighbors have an agent positioned that was initially singleton, then $r_u$ becomes a local leader if $u$ has the smallest degree among the neighbors (Lines 13 and 14 of Algorithm \ref{algorithm:local_leader_election}). 
%
If there is at least one same degree neighbor, $u$ has to have a higher ID agent positioned than the one positioned at that same degree neighbor to become a local leader, otherwise  $r_u$ becomes $non\_candidate$ (Lines 5 and 6 of Algorithm \ref{algorithm:local_leader_election}).  

If not all neighbors have agent positioned and $r_u$ does not have a neighbor of higher degree and does not find any agent belonging to $Multiplicity\_Election$ or status local leader, then $r_u$ explores neighbors after $2\delta_v$ rounds if $\delta_u<\delta_v$, where $v$ is a neighboring node. $\forall$ neighbor $v, \delta_v>\delta_u$ case simply demands visiting the neighbors after $2\delta_v$ rounds. In $2\delta_v$ rounds, if there was an agent $r_v$ at $v$ initially then $r_v$ will become the $non\_candidate$ agent after finding the $\delta_u<\delta_v$ (Lines 9-11 of Algorithm \ref{algorithm:local_leader_election}). Otherwise, if there is no such agent initially, then the agent at node $v$ would come from $Multiplicity\_Election$. This ensures that there exists a $local\_leader$, therefore, $r_u$ becomes $non\_candidate$ with or without finding $r_v$ after $2\delta_v$ rounds. 
For $\delta_v=\delta_u$, we need a guarantee that $r_u$ meets $r_v$ if an initially singleton agent $r_v$ is present at $v$. This is a challenging situation in Algorithm \ref{algorithm:local_leader_election} ($Singleton\_Election$) which is handled through Algorithm \ref{algorithm:padding} ($Neighbor\_Exploration\_with\_Padding$) providing a guarantee that $r_u$ meets $r_v$ and vice-versa (Lines 7 and 8 of Algorithm \ref{algorithm:local_leader_election}).

Algorithm \ref{algorithm:padding} ($Neighbor\_Exploration\_with\_Padding$) works as follows.
Suppose an agent $r_u$ has an ID of $b$ bits; note that $b\leq c\cdot \log n$ for some constant $c>1$. We pad a sequence of `10' bits $b^2$ times 
to the LSB (least significant bit) of the $r_u$'s ID, i.e., 
\begin{eqnarray*}
b\underbrace{10}_{\mbox{$1$}}\underbrace{10}_{\mbox{$2$}}\ldots\underbrace{10}_{\mbox{$b^2$}}.
\end{eqnarray*}

Now the ID of $b$ bits becomes the ID of $b+2b^2$ bits. 
This padding will be helpful in making the same degree neighboring agents meet each other. 
Algorithm \ref{algorithm:padding} ($Neighbor\_Exploration\_with\_Padding$) ran by agent $r_u$ starts from its MSB (most significant bit) and ends at LSB. If a bit is 1, then $r_u$ explores all neighbors which finishes in $2 \delta_u$ rounds, however, when a bit is $0$, it remains at its position for $2\delta_u$ rounds. 
We will show that using this padding approach the agents at the same degree neighbors meet each other in $O(\delta_u\log^2n)$ rounds, a crucial component in Algorithm \ref{algorithm:local_leader_election} ($Singleton\_Election$). 

\vspace{2mm}
\noindent{\bf Multiplicity\_Election (Algorithm \ref{algorithm:dispersion}).}
This procedure is run by agents which were initially non-singleton. Let $u$ be a node where $r_u$ is positioned and it is the minimum ID agent among the co-located (say, group-$r_u$). $r_u$ is responsible for running $Multiplicity\_Election$, others follow $r_u$. $Multiplicity\_Election$ is essentially a DFS traversal procedure (denoted as $DFS(r_u)$) to visit the nodes of $G$ in forward and backtrack phases \cite{KshemkalyaniMS20}. At $u$, the largest ID agent in group-$r_u$ settles and it tracks parent and child pointers for the $DFS(r_u)$. Each node visited by $DFS(r_u)$ may be {\em occupied} (has an agent positioned) or empty. If an agent is positioned on a newly visited node, it writes ID and parent and child pointer information about $DFS(r_u)$ and continues. If an empty node, it positions the largest ID agent in group-$r_u$ and continues. At some point in time, $DFS(r_u)$ reaches an empty node such that $r_u$ is the only agent on it and the parent node of $r_u$ in $DFS(r_u)$ is non-empty. $r_u$ elects itself as a local leader and $Multiplicity\_Election$ is finished for $r_u$. If $r_u$ finds the parent node of $r_u$ in $DFS(r_u)$ is empty when becoming singleton, it waits to either become a local leader or non-candidate by going to the parent in $DFS(r_u)$ and waiting there until it meets the agent which has this parent node as its home node (Line 2 of Algorithm \ref{algorithm:dispersion}).

While running $DFS(r_u)$, we call the node where $r_u$ is currently positioned the {\em head} node of $DFS(r_u)$.  $DFS(r_u)$ may meet $DFS(r_v)$ from another agent $r_v$. 
We differentiate two cases.
If the meeting happens at a node $w$ such that $w$ is the head node of both $DFS(r_u)$ and $DFS(r_v)$ and $w$ is an empty node, then an agent from $DFS(r_u)$ settles if $r_u<r_v$, otherwise an agent from $DFS(r_v)$ settles at $w$. Both write their DFS information on the agent positioned on $w$ and continue their traversal. If $w$ is a non-empty node, they both continue without settling any agent but just write their DFS information at the agent positioned on $w$. 
However, if $DFS(r_u)$ meets $DFS(r_v)$ at a node $w$ which is not a head node of $DFS(r_v)$ ($w$ is a head node of $DFS(r_u)$ since $DFS(r_u)$ is meeting $DFS(r_v)$ at $w$), then $DFS(r_u)$ simply continues writing its DFS information on the agent positioned at $w$. 

To deal with the situation that $DFS(r_u)$ does not miss meeting an agent doing $Singleton\_Election$, $DFS(r_u)$ waits at every node it visits for a round before exiting.  This is enough since an agent doing $Singleton\_Election$ returns to its node every second round. If the node is still empty after the (waited) round, it runs $Confirm\_Empty()$ (Algorithm \ref{algorithm:confirm-empty}) to verify whether it is a (possible) home node of a local leader (an agent waiting to become a local leader, see Remark in Algorithm \ref{algorithm:dispersion}), or indeed an empty node.

\vspace{2mm}
\noindent{\bf Global\_Election (Algorithm \ref{algorithm:leader-election}).}
This procedure is run by agents who become local leaders. Notice that an agent may become a local leader running either Algorithm \ref{algorithm:local_leader_election} ($Singleton\_Election$) or Algorithm \ref{algorithm:dispersion} ($Multiplicity\_Election$). As soon as an agent becomes a local leader, it contends to become a unique global leader. Let $r_u$ becomes a local leader at node $u$ ($u$ being the home node of $r_u$) at round $roundNo_u$. It then starts a DFS traversal denoted by tuple $DFS(roundNo_u,r_u)$, the goal of which is to see whether it can visit all the edges of $G$ and return to $u$, its home node. If $r_u$ can do so, it declares itself as a global leader. While running $DFS(roundNo_u,r_u)$, it might meet $DFS(roundNo_v,r_v)$. $DFS(roundNo_u,r_u)$ continues if its ID is lexicographically larger than $DFS(roundNo_v,r_v)$, otherwise it stops and becomes a $non\_candidate$ and returns to node $u$ (its home node). Additionally,  $DFS(roundNo_u,r_u)$ might meet $Multiplicity\_Election$ and if the meeting happens at the head node of $Multiplicity\_Election$, it stops, otherwise it continues. Additionally, if $DFS(roundNo_u,r_u)$ reaches a confirmed empty node (i.e., it is not an empty home node) then it also stops as it knows that $Multiplicity\_Election$ from some other initially non-singleton node is still going on since the node is still empty. 

\subsection{Analysis of the Algorithm}
We analyze Algorithm \ref{algorithm:leader_election} for correctness and its complexity guarantees. We start with $Singleton\_Election$ (Algorithm~\ref{algorithm:local_leader_election}).

\begin{lemma}\label{lem:rounds_with_same_degree_leader}
    In Singleton\_Election (Algorithm~\ref{algorithm:local_leader_election}) run by agent $r_u$ at node $u$, if there is a neighboring agent $r_v$ positioned on the neighbor node $v$ such that $\delta_u=\delta_v$ and $\delta_u,\delta_v$ both being the minimum, $r_u$ meet $r_v$ in $O(\delta_u \log^2 n)$ rounds, running  $Neighbor\_Exploration\_with\_Padding$ (Algorithm \ref{algorithm:padding}).
\end{lemma}
\begin{proof}
Since agent IDs are from the interval $[1, n^{O(1)}]$, each agent has ID of size $\leq c\cdot \log n$ for some constant $c$. Therefore, for any two agents $r_u,r_v$, two cases exist for the number of bits on their IDs: either equal or unequal. Consider the equal case first. If $r_u$ and $r_v$ have an equal number of bits (say, $b$) then their IDs must be different in at least one bit since IDs are unique, i.e., if one has bit '1' at $\beta$-th place from MSB, another must have bit '0' at $\beta$-th place from MSB. Since an agent explores neighbors while the bit is '1' and stays at its place when the bit is '0', $r_u$ finds $r_v$ within $2\cdot \delta_{u} \cdot b$ rounds. Since the bit may be different at the $b$-th place from MSB, the total time taken for $r_u$ to meet $r_v$ is $O(\delta_{u} \log n)$ rounds.

Now consider the case of an unequal number of bits. Let $r_u$ and $r_v$, respectively, have  $b$ and $d$ bits in their IDs with $b\neq d$. W.l.o.g., $b>d$, i.e., $b = d+c_1$, where $d, c_1 \geq 1$. Therefore, the total number of bits after padding in the agent $r_u$'s ID is $b+2b^2$. 
 We have that $$b+2b^2=(d+c_1) + 2(d+c_1)^2 = 2d^2+2c_1^2+4 \cdot d \cdot c_1+d+c_1.$$ Similarly, the total number of bits in the agent $r_v$ after padding is $d+2d^2$. Therefore, after padding, the difference in the number of bits of the IDs of $r_u$ and $r_v$ is $$2c_1^2+4 \cdot d \cdot c_1+c_1.$$ 
 Since $d,c_1\geq 1$, the difference is at least $7$ bits in the overall length of the IDs of $r_u$ and $r_v$ after padding. Additionally, out of these $7$ bits, at least  $3$ bits are `1's during which agent $r_u$ can explore the node $v$ with $r_v$ positioned at node $v$. 
    What that means is, if $r_v$ (the smaller ID than $r_u$) stops after $\delta_{v}(d+2d^2)$ rounds, then there are at least $3$ chances for $r_u$ to meet $r_v$ at its node $v$ since $r_u$ with bit `1' will be visiting its neighbors and $r_v$ is at $v$ not moving anymore since it finished visiting its neighbors. Therefore, the round complexity becomes   $O(\delta_{u} (b+2b^2))=O(\delta_u \log ^2n)$ rounds, since $b\leq c\cdot \log n$. 
\end{proof}

\begin{lemma}\label{lem: rounds_lesser_than_O(m)}
    In Lemma \ref{lem:rounds_with_same_degree_leader}, $O(\delta_u\log^2n) < O(m)$. 
    \end{lemma}
\begin{proof}
    Notice that $\delta_u=\delta_v$ and $\delta_u$ is the minimum among the neighbor degrees of node $u$. Therefore, it must be the case that there are at least $\delta_{u}(\delta_{u}-1)/2$ edges in $G$. We now relate this to $m$.  
    We consider two cases 
    and show that in both the cases $O(\delta_u\log^2n) < O(m)$.
    
\begin{itemize}
    \item $\delta_{u}(\delta_{u}-1)/2 \leq O(\delta_{u} \log^2 n)$. This implies that $\delta_u\leq O(\log^2 n)$. Which means $$\delta_{u}(\delta_{u}-1)/2 \leq O(\log^4n)<O(n)<O(m).$$
   
    \item $\delta_{u}(\delta_{u}-1)/2 > O(\delta_{u} \log^2 n).$ This implies that $m\geq \delta_{u}(\delta_{u}-1)/2>O(\delta_{u} \log^2 n).$ Which means $$O(m)>O(\delta_{u} \log^2 n).$$
\end{itemize}    
Therefore, in Lemma \ref{lem:rounds_with_same_degree_leader}, $O(\delta_{u} \log^2 n)<O(m)$. 
\end{proof}

\begin{lemma}
\label{lemma:local-leader-election}
  An initially singleton agent $r_u$ at node $u$ running  Singleton\_Election (Algorithm~\ref{algorithm:local_leader_election}) either becomes a local leader or a non-candidate within $O(m)$ rounds.   
\end{lemma}
\begin{proof}
    We consider two cases: (A) all $\delta_u$ neighbors of node $u$ where $r_u$ is positioned have agents positioned initially,
    (B) at least 1 neighbor of $u$ was initially empty.

 We first consider Case A. We have two sub-cases: (A1) all $\delta_u$ neighbors have a singleton agent initially, (A2) at least a neighbor has multiple agents initially. 
 In Case A1, we prove that $r_u$ either becomes a local leader or non-candidate in $O(\delta_u \log^2 n + \max_{v \in N(r_u)} \delta_v)$ rounds. The proof is as follows. If there is a neighbor node $v$, such that $\delta_u>\delta_v$, then $r_u$ becomes a non-candidate in $2\delta_u$ rounds. If $u$ is the smallest degree node in $N(r_u)\cup\{u\}$  and there is a node $v'\in N(r_u)$ with $\delta_u=\delta_{v'}$ then either $u$ becomes a local leader or a non-candidate in $O(\delta_u\log^2n)$ rounds (Lemma \ref{lem:rounds_with_same_degree_leader}).  If $u$ is the smallest degree node in $\{u\}\cup N(r_u)$ and there is no neighbor $v'$ with $\delta_u=\delta_{v'}$, then to become a local leader, it needs to know that the largest degree neighbor, say $v$, in fact, has an agent positioned. $r_u$ gets to know there is an agent $r_v$ at node $v$ when $r_v$ finishes visiting all its neighbors which takes $2\delta_v$ rounds. After that $r_v$ becomes non-candidate (since it knows of $u$ with $\delta_u<\delta_v$) and stays at $v$. 
 Therefore, combining the times for all the above cases, we have the time complexity for Case A1 $O(\delta_u \log^2 n + \max_{v \in N(r_u)} \delta_v)$ rounds.

 In Case A2, $r_u$ gets to know the neighbor(s) with multiple agents is going to run $Multiplicity\_Election$ (Algorithm \ref{algorithm:dispersion}) and become a local leader. This information can be collected by $r_u$ within $2\delta_u$ rounds since one agent at the multiplicity neighbor node does not move during Algorithm \ref{algorithm:leader_election}.

 We now consider Case B. We have two sub-cases: (B1) all the neighbors that have agents positioned are all singleton agents, (B2) at least one neighbor has multiple agents. In Case B1, $r_u$ cannot become a local leader in $O(\delta_u \log^2 n + \max_{v \in N(r_u)} \delta_v)$ rounds if it does not meet any agent at (at least) a neighbor in those $O(\delta_u \log^2 n + \max_{v \in N(r_u)} \delta_v)$ rounds. $r_u$ remains at $u$ as a candidate until it gets to know of $Multiplicity\_Election$ or $Global\_Election$ and when it knows of one such procedure, it becomes a non-candidate. Since $Multiplicity\_Election$ and  $Global\_Election$ finish in $O(m)$ time, $r_u$ becomes a non-candidate in $O(m)$ rounds (Lemmas \ref{lemma:dispersion} and \ref{lemma: leader-election}). However, if it meets an agent within $O(\delta_u \log^2 n + \max_{v \in N(r_u)} \delta_v)$ rounds, then that must be from $Multiplicity\_Election$ since a neighbor was initially empty, and hence $r_u$ becomes non-candidate in $O(\delta_u \log^2 n + \max_{v \in N(r_u)} \delta_v)$ rounds.  
 In Case B2, $r_u$ knows it cannot become a local leader in $2\delta_u$ round as in Case A2. 
 Therefore, the total time for both cases A and B is $O(m)$ rounds.
 \end{proof}

For the dispersed initial configurations, we prove below that at least one agent at a node (despite being singleton initially) becomes a local leader.

 \begin{lemma}
     \label{lemma:a-local-leader}
Starting from a dispersed initial configuration, at least one agent becomes a local leader running Algorithm \ref{algorithm:local_leader_election} ($Singleton\_Election$).     
 \end{lemma}
 \begin{proof}
    Let node $u$ be the smallest degree node in $G$, i.e., $\delta_u=\min_{v\in G}\delta_v$. 
    We have two cases: (i) $u$ is the unique smallest degree node in $G$, i.e., there is no other node $v'$ such that $\delta_u=\delta_{v'}$ (ii) there is at least a node $v'\in G$ with $\delta_{v'}=\delta_u$.
    We first consider Case (i). Since $u$ is the unique smallest degree node, $r_u$ meets all its neighbors in $O(\delta_u \log^2 n + \max_{v \in N(r_u)} \delta_v)$ rounds (Case A1 in Lemma \ref{lemma:local-leader-election}). Since each neighbor of $u$ is of a higher degree than $u$, they all become non-candidate and $r_u$ becomes a local leader. Now consider Case (ii). We have two sub-cases: (ii.A) $u$ and $v'$ are neighbors (ii.B) $u$ and $v'$ are not neighbors. In Case (ii.B) $r_u$ is elected as a local leader as discussed in Case (i). For Case (ii.A), we have from Lemma \ref{lem:rounds_with_same_degree_leader} that $r_u$ meets $r_{v'}$ in $O(\delta_u\log^2n)$ rounds. After that, either $r_u$ or $r_{v'}$ becomes a non-candidate. If $r_{v'}$ becomes non-candidate, $r_u$ becomes a local leader and we are done. Otherwise, if $r_{v'}$ becomes a non-candidate, then it has a neighbor $v''$ with $\delta_{v''}=\delta_{v'}=\delta_u$ and $r_{v''}$ remains as a candidate to become a local leader. This chain stops at the first node $v*$ with the same degree neighbor $v'''$ such that $r_{v'''}.ID< r_{v*}.ID>\ldots>r_{v''}.ID>r_{v'}.ID>r_u.ID$ and $v*$ becomes a local leader.  
 \end{proof}

We now consider rooted and general initial configurations. 

\begin{lemma}
\label{lemma:dispersion}
    $Multiplicity\_Election$ (Algorithm~\ref{algorithm:dispersion}) run by an initially non-singleton agent $r_u$ of minimum ID among the $x>1$ co-located agents finishes positioning those co-located agents on $x$ different nodes of $G$ in $O(m)$ rounds with $O(n\log n)$ bits memory per agent. 
\end{lemma}
\begin{proof}
Consider first the rooted initial configuration. Since there is no singleton agent initially, $Singleton\_Election$ does not run. We have from Kshemkalyani {\it et al.} \cite{KshemkalyaniMS20} that the DFS traversal $DFS(r_u)$ run by the minimum ID agent finishes achieving a dispersed configuration in $\min(4m-2n+2,4n\Delta)=4m-2n+2=O(m)$ rounds using $O(\log(n+\Delta))=O(\log n)$ bits per agent, since $m\leq n\Delta$ and $\Delta\leq n$. 

Now consider the general initial configurations of $\ell$ DFSs ran by $\ell$ minimum ID agents from $\ell$ non-singleton nodes of $G$. Notice that $\ell\leq n/2$. 
We have two situations: 
\begin{itemize}
    \item [(i)] there is no  singleton node initially, i.e. $n$ agents are on $\ell$ non-singleton nodes 
    \item [(ii)] there is at least a singleton node, i.e.,  $n$ agents are on $\ell$ non-singleton nodes and at least one singleton node.    
\end{itemize}
Consider the first situation. 
Since there is no singleton node initially, $Singleton\_Election$ does not run and hence the synchronization is only between $Multiplicity\_Election$ procedures. 
A DFS with $x$ initially co-located agents needs to visit $x-1$ other empty nodes to settle all its co-located agents. When an empty node is visited by a single DFS, then an agent from it settles. If an empty node is visited by the heads of two or more DFSs, an agent from one DFS settles. 
Therefore, since there are $\ell$ DFSs with $n-\ell$ empty nodes and $n-\ell$ agents to find the empty nodes to settle and in each empty node visited by one or more DFSs an agent settles, a DFS may need to traverse all the edges of $G$ to be able to settle all its agents at empty nodes. We know that traversing all the edges of $G$ finishes in $4m-2n+2$ rounds since each DFS continues until it settles all its agents. 
Regarding memory, since each DFS continues until it is able to settle all its agents, an agent positioned at a node may need to store the information about all the $\ell$ DFSs. Since $\ell\leq n/2$ and for a DFS a node needs to store $O(\log n)$ bits, the total memory needed at an agent is $O(n\log n)$ bits.  

Now consider the second situation, i.e., there is at least an agent that runs $Singleton\_Election$.  If a DFS visits an empty node, say $u$, then it has to confirm whether: 
\begin{itemize}
    \item [(i)] it is in fact an empty node, 
    \item [(ii)] a node of agent running $Singleton\_Election$, \item [(iii)] a home node of an agent running $Global\_Election$ after becoming a local leader through $Singleton\_Election$, or 
    \item [(iv)] a home node of agent running $Global\_Election$ after becoming a local leader through $Multiplicity\_Election$. 
\end{itemize}
In the first case, the waiting for a round results $u$ to be empty. The DFS then runs procedure $Confirm\_Empty()$ (Algorithm \ref{algorithm:confirm-empty}) which confirms it to be indeed an empty node. 
The second case is confirmed since DFS waits at a node for a round and an agent running $Singleton\_Election$ returns to its node every two rounds. For the third case, the head of DFS needs to visit the port-1 neighbor of $u$ to find out whether it is a home node of the local leader. The fourth case demands the head of DFS to visit all the neighbors of that node. Therefore, for each empty node $u$ reached, this confirmation needs at most $2\delta_u+1$ rounds, $2\delta_u$ rounds to visit all neighbors running $Confirm\_Empty()$ (Algorithm \ref{algorithm:confirm-empty}), and $1$ round wait at $u$ by DFS before running $Confirm\_Empty()$ (Algorithm \ref{algorithm:confirm-empty}). After a home node (which is empty) is confirmed then the agent running DFS can store this information for future use. That is, if the agent running the DFS visits $u$ again and finds it to be empty, it can simply use the stored information to decide whether it is a home node. Since there are $n$ nodes, total $O(n\log n)$ bits is enough for an agent running DFS to store this home node information for future use. Therefore, $Multiplicity\_Election$ has the additional overhead of at most $O\big(\sum_{i=1}^n \delta_i \big)=O(m)$ due to $Confirm\_Empty$ (Algorithm \ref{algorithm:confirm-empty}). 

Combining the above costs, $Multiplicity\_Election$ (Algorithm \ref{algorithm:dispersion}) for each initially non-singleton agent $r_u$ finishes in $O(m)$ rounds with memory per agent $O(n\log n)$ bits. 
\end{proof}


\begin{lemma}
\label{lemma:dispersion-election}
    Suppose there were $\ell\geq 1$ multiplicity nodes in the initial configuration. 
    \begin{itemize}
        \item 
    If $\ell=1$, an initially non-singleton agent $r_u$ of minimum ID becomes a local leader running Algorithm~\ref{algorithm:dispersion} ($Multiplicity\_Election)$. 
    \item If $\ell\geq 2$, at least 2  initially non-singleton agents of minimum ID among the $x>1$ co-located agents in their respective multiplicity  nodes become local leaders running Algorithm~\ref{algorithm:dispersion}. 
    \end{itemize}
\end{lemma}
\begin{proof}
Consider first the case of $\ell=1$. There is a single procedure $Multiplicity\_Election$ (Algorithm \ref{algorithm:dispersion}) running. When $Multiplicity\_Election$ (Algorithm \ref{algorithm:dispersion}) finishes, let $w$ be the node on which the initially non-singleton agent $r_u$ of minimum ID is positioned. Let $T_u$ be the DFS tree built during $Multiplicity\_Election$ (Algorithm \ref{algorithm:dispersion}). Let $w'$ be the parent node of $w$ in $T_u$. Note that $w'$ has the initially non-singleton agent of second minimum ID $r_{w'}$ positioned. Therefore, $r_u$ can become a local leader writing its home node information at the agent $r_{w'}$ positioned at $w'$. 

Consider now the case of $\ell\geq 2$. There will be $\ell$ instances of $Multiplicity\_Election$ (Algorithm \ref{algorithm:dispersion}) running possibly concurrently. When a $Multiplicity\_Election$ (Algorithm \ref{algorithm:dispersion}) instance finishes, let $w$ be the node on which it finishes.  Notice that, the initially non-singleton agent $r'$ of minimum ID from that finished instance is positioned on $w$. Let $T_{r'}$ be the DFS tree built by agent $r'$ while running $Multiplicity\_Election$. Let $w'$ be the parent node of $w$ in $T_{r'}$. If $w'$ is non-empty, $r_u$ can become a local leader writing its home node information at the agent $r''$ positioned at $w'$. Therefore, if non-empty parent condition in $T_{r'}$  satisfies for each $Multiplicity\_Election$ (Algorithm \ref{algorithm:dispersion}) instance, there will be $\ell$ local leaders. 

Suppose non-empty parent condition in $T_{r'}$  does not satisfy for $r'$. For this to happen, another instance of $Multiplicity\_Election$ by some agent $r''$ must have previously finished on the parent node $w'$ of $r'$ in $T_{r'}$. and $r''$ must have become a  local leader at $w'$. For $r'$ to find $w'$ empty, $r''$ must have now running $Global\_Election$ (Algorithm \ref{algorithm:leader-election}). Since $w'$ is empty, $r'$ cannot write its home node information at $w'$ and hence cannot immediately become a local leader.

The agent $r'$ now comes to $w'$ and waits until $r{''}$ returns to $w'$. During the wait, no other agent $r'''$ occupies node $w$, since when $r'''$ runs $Confirm\_Empty()$, it finds that $r'$ which was supposed to be at $w$ is waiting at the parent node $w'$ of $T'$. $r'''$ then tries to position itself at a neighbor, say $w''$, of $w$.  If $w''$ is empty, then the parent node in $T'''$ is $w$ which is empty (since $r'$ is waiting at $w'$). This process forms a chain of nodes (i.e., each node in the chain being the parent in the DFS tree of an agent) such that the first node in the chain is the home node of a local leader and all other nodes in the chain are the home nodes of the agents waiting (on the predecessor nodes in the chain) to become local leaders. After the local leader  returns to $w'$, the waiting agents in the chain go back to their nodes (traversing the child pointers), i.e., $r'$ waiting at $w'$ goes to $w$, $r'''$ waiting at $w$ goes to $w''$, and so on. 
When a waiting agent $r_1$ goes to its home node $h_1$ and finds a waiting agent $r_2$ positioned at $h_1$, then $r_1$ knows that $r_2$ must have started waiting later in time compared to $r_1$. Therefore, $r_1$ becomes a non-candidate and stays at $h_1$. $r_2$ goes to its home node, say $h_2$, and finds an agent $r_3$ waiting, then $r_2$ becomes a non-candidate and stays at $h_2$, and the process repeats. Otherwise, $r_2$ knows that it is the last waiting agent in the chain. $r_2$ becomes a local leader at $h_2$ and writes the home node information at agent $r_1$ at node $h_1$, which is the parent node in the DFS tree of $r_2$. Therefore,  at least two initially non-singleton agents become local leaders during  $Multiplicity\_Election$ (Algorithm \ref{algorithm:dispersion}). 
\end{proof}

\begin{lemma}
\label{lemma:dispersion-election-time}
  In Lemma \ref{lemma:dispersion-election}, each initially non-singleton agent $r_u$ running $Multiplicity\_Election$ 
  either becomes a local leader or a non-candidate in $O(m)$ rounds.   
\end{lemma}
\begin{proof}
We have from Lemma \ref{lemma:dispersion}, agents achieve a dispersed configuration in $O(m)$ rounds, if they were not initially in a dispersed configuration. Therefore, for $\ell=1$ in Lemma \ref{lemma:dispersion-election}, an initially non-singleton agent of minimum ID $r_u$ in the single multiplicity node becomes a local leader in $O(m)$ rounds. For $\ell\geq 2$, if non-empty parent condition is satisfied for each of $\ell$ initially non-singleton agents of minimum ID in their respective multiplicity nodes, then they all become local leaders in $O(m)$ rounds because they do not wait. 
If non-empty parent condition is not satisfied, then for each chain of waiting agents, one agent in the parent node in the DFS tree of the first waiting agent in the chain becomes a local leader in $O(m)$ rounds since it does not wait. That local leader in the chain running $Globasl\_Election$ (Algorithm \ref{algorithm:leader-election}) must return to its home node in the next $O(m)$ rounds, since it runs a DFS traversal which must finish in $O(m)$ rounds \cite{Kshemkalyani}. After that, if the chain is of length $\ell'\leq \ell$ then the last waiting agent in the chain reaches its node in $O(\ell')$ rounds and becomes a local leader. Other waiting agents in the chain become non-candidates. Since $\ell'<\ell\leq n/2$, we have total time $O(m+\ell')=O(m)$ rounds.  
\end{proof}

\begin{lemma}
\label{lemma:occupyhomenode}
Consider a currently empty home node $u$ of a local leader agent $r_u$ running $Global\_Election$ (Algorithm \ref{algorithm:leader-election}). 
\begin{itemize}
\item If $r_u$ became a local leader through $Singleton\_Election$ (Algorithm \ref{algorithm:local_leader_election}), 
node $u$ will not be occupied by an agent other than $r_u$.
\item If $r_u$ became a local leader through $Multiplicity\_Election$ (Algorithm \ref{algorithm:dispersion}), 
node $u$ may be occupied for a short period by an agent waiting to become a local leader at the neighbor of $u$ until $r_u$ returns to $u$. 
\end{itemize}
\end{lemma}
\begin{proof}
We prove the first case.
Notice that when an agent becomes a local leader at node $u$ through from $Singleton\_Election$ (Algorithm \ref{algorithm:local_leader_election}), then all the neighbors must have an initially non-singleton agent positioned. 
Therefore, when $r_u$ becomes a local leader at $u$ from $Singleton\_Election$ (Algorithm \ref{algorithm:local_leader_election}), then the information that $u$ is a home node is written in port-1 neighbor of $u$ by $r_u$ before it initiates  $Global\_Election$ (Algorithm \ref{algorithm:leader-election}) as that port-1 neighbor is non-empty and not moving. For an agent to settle at $u$ when it finds empty, it has to confirm that it is in fact empty, i.e., not a home node of another agent. When the agent runs $Confirm\_Empty$ (Algorithm \ref{algorithm:confirm-empty}), it finds that $u$ is in fact a home node. Therefore, no other agent occupies $u$, the home node of local leader agent $r_u$. 

We now prove the second case. Notice that when an agent $r_u$ becomes a local leader at node $u$ through from $Multiplicity\_Election$ (Algorithm \ref{algorithm:dispersion}), there is no guarantee that all neighbors of $u$ have agents positioned except the parent node in the DFS tree of $r_u$ at $u$. When an agent $r*$ finishes its $Multiplicity\_Election$ (Algorithm \ref{algorithm:dispersion}) instance at the empty-neighbor, say $w''$, of $u$, then $r*$ finds that $u$ (the parent node in its DFS tree) empty and has to wait to become a local leader or non-candidate. $r*$ waits at $u$ until $r_u$ running $Global\_Election$ (Algorithm \ref{algorithm:leader-election}) returns to $u$. After that $r*$ leaves $u$ and enters $w''$ at which it becomes either a local leader or non-candidate. 
\end{proof}

\begin{lemma}
\label{lemma:return-home-node}
Consider a local leader agent $r_u$ running $Global\_Election$ (Algorithm \ref{algorithm:leader-election}). As soon as $r_u$ realizes it cannot become a global leader, it can return to its home node $u$.  
\end{lemma} 
\begin{proof}
Recall that $r_u$ runs a DFS traversal $DFS(roundNo_u,r_u)$ during $Global\_Election$ (Algorithm \ref{algorithm:leader-election}). $DFS(roundNo_u,r_u)$ builds a DFS tree $T$ with its root node the home node $u$ of $r_u$. In $T$, there is a sequence of parent pointers from the current node position (which is the head of $DFS(roundNo_u,r_u)$) of $r_u$ to the root node of $T$. Since every local leader runs its separate DFS traversal and maintains the tree information in each node its DFS visits, $r_u$ can follow the parent pointers in $T$ until reaching the root node, which is its home node.  
\end{proof}

\begin{lemma}
\label{lemma: leader-election}
Global\_Election (Algorithm \ref{algorithm:leader-election}) elects a unique global leader and terminates in $O(m)$ rounds with $O(\log n)$ bits at each agent. 
\end{lemma}
\begin{proof}
An agent that is not a local leader does not run $Global\_Election$ (Algorithm \ref{algorithm:leader-election}). We have from Lemmas \ref{lemma:a-local-leader} and  \ref{lemma:dispersion-election} that at least one agent becomes a local leader starting from any initial configuration. Consider an agent $r_u$ that becomes a local leader at $roundNo_u$. It initiates $DFS(roundNo_u,r_u)$ at $roundNo_u$. 
Suppose it visits an empty node $w$ at some round $t>roundNo_u$. We have from the proof of Lemma \ref{lemma:dispersion} that it can be confirmed in $O(\delta_w)$ rounds that whether $w$ is in fact empty or a home node running $Confirm\_Empty$ (Algorithm \ref{algorithm:confirm-empty}).  If it is empty, there must be at least one $Multiplicity\_Election$ still running and an agent will become a local leader in some round $t'>t$, and hence $DFS(roundNo_u,r_u)$ stops. If $DFS(roundNo_u,r_u)$ does not visit any empty node, then it must visit another $DFS(roundNo_v,r_v)$. We have that either $DFS(roundNo_u,r_u)$ stops or $DFS(roundNo_v,r_v)$ stops due to this meet but not both. Since the IDs are unique, there is always one and only one $DFS(roundNo_w,r_w)$ run by local leader $w$ that wins all the competitions on the DFSs. Therefore, we have a unique leader.  
Since $Confirm\_Empty$ (Algorithm \ref{algorithm:confirm-empty}) needs to be run for $n$ different nodes at most once with total overhead of $O\big(\sum_{i=1}^n \delta_i \big)=O(m)$ rounds and each DFS takes $O(m)$ rounds,  the total time to finish $Global\_Election$ (Algorithm \ref{algorithm:leader-election}) after $Multiplicity\_Election$ is $O(m)$ rounds. To store information about $Global\_Election$ run by all local leaders, $O(\log n)$ bits at each agent is sufficient since the the DFSs synchronize on which one stops and which one proceeds based on their DFS IDs. 
\end{proof}

\noindent{\bf Proof of Theorem \ref{theorem:leader}:}
\begin{proof}
    In the rooted initial configuration only one $Multiplicity\_Election$ (Algorithm \ref{algorithm:dispersion}) instance runs until $n$ agents disperse to $n$ nodes and the minimum ID agent running $Multiplicity\_Election$ becomes a local leader. It then runs $Global\_Election$ (Algorithm \ref{algorithm:leader-election}) and becomes a global leader. The total time will be $O(m)$ rounds (Lemmas \ref{lemma:dispersion} and \ref{lemma: leader-election}). 

    In the dispersed initial configuration, only $Singleton\_Election$ (Algorithm \ref{algorithm:local_leader_election}) runs. The agents either become local leaders or non-candidates in $O(m)$ rounds (Lemma \ref{lemma:local-leader-election}) and there will be at least one agent elected as a local leader (Lemma \ref{lemma:a-local-leader}). $Global\_Election$ then runs for the next $O(m)$ rounds for an agent among local leaders to become a global leader.

    For the case of $\ell$ non-singleton nodes and no singleton node initially, again, $\ell$ instances of $Multiplicity\_Election$ finish in $O(m)$ time electing at least 2 local leaders and they again synchronize in the lexicographical order of the DFS IDs while running $Global\_Election$ so that one local leader agent becomes a global leader in next $O(m)$ rounds (Lemmas \ref{lemma:dispersion}, \ref{lemma:dispersion-election-time}, and \ref{lemma: leader-election}). 

    Now the only remaining case is the mix of singleton and non-singleton nodes. In this case, $Multiplicity\_Election$ finishes in $O(m)$ rounds electing at least one agent as a local leader for any $\ell\geq 1$ (Lemmas \ref{lemma:dispersion-election} and \ref{lemma:dispersion-election-time}) and by that time any agent running $Singleton\_Election$ either becomes a local leader or gets to know $Multiplicity\_Election$ and stops running $Singleton\_Election$ changing its status to ``non\_candidate''. After that $Global\_Election$ finishes in $O(m)$ rounds electing one agent that becomes a local leader as a global leader. Therefore, the total runtime is $O(m)$ rounds.

    Consider the local leaders (through both $Singleton\_Election$ and $Multiplicity\_Election$) that cannot become a global leader. We have from Lemmas \ref{lemma:occupyhomenode} and \ref{lemma:return-home-node} that they can return to their home nodes and stay at those nodes as non-candidates. Notice that returning to home nodes takes $O(n)$ rounds since the agents can traverse parent pointers on the DFS tree they build while running $Multiplicity\_Election$.   

    Regarding memory, $Multiplicity\_Election$ and $Global\_Election$ need to run $Confirm\_Empty()$ (Algorithm \ref{algorithm:confirm-empty}) and store the information. Additionally, the $\ell$ DFSs during $Multiplicity\_Election$ and all local leader DFSs during $Global\_Election$ ask nodes to keep their information; hence, $O(n\log n)$ bits per agent are needed. All the other variables are of either $O(1)$ size or $O(\log n)$ size and there are only a constant number of them. For the dispersed initial configuration, $Global\_Election$ can be carried out with only $O(\log n)$ bits per agent and  $Singleton\_Election$ needs $O(\log^2n)$ bits per agent, and hence total $O(\log^2n)$ bits per agent.  
\end{proof}

\section{MST Construction}\label{sec: MST construction}
In this section, we present a deterministic algorithm to construct an MST of $G$ given a leader $r_l$ elected in the previous section and its DFS tree $T_{r_l}$ built while running $DFS(roundNo_l,r_l)$. The MST construction finishes in $O(m)$ rounds with $O(\Delta \log n)$ bits at each agent. We consider $G$ to be weighted and hence our MST is a minimum weight MST of $G$. 

\vspace{2mm}
\noindent{\bf Overview of the Algorithm.}
Starting from any arbitrary initial configuration $n$ agents, when Algorithm \ref{algorithm:leader_election} finishes electing a leader, $n-1$ nodes of $G$ have an agent each positioned with status $non\_candidate$ and one node has an agent with status $leader$. 

Let $r_v$ be a leader positioned at node $v$. 
We can have two methods for MST construction. The first method is to ask the leader $r_v$ to collect all the agents at node $v$, making a rooted configuration. This can be done by revisiting the DFS tree $T_{r_l}$ built by $r_v$ while running $DFS(roundNo_v,r_v)$ to elect itself as a leader during Algorithm \ref{algorithm:leader-election} ($Global\_Election$), collecting the agents at node $v$ in $O(n)$ rounds. 
After that $r_v$ can run $DFS(r_v)$ as in Algorithm \ref{algorithm:dispersion} to disperse the agents as well as assign ranks $1$ to $n$. The leader will have rank $1$ and the agent that settles $i$-th in the DFS order of empty node visits will have rank $i$.
The second method is to run $DFS(r_v)$ to visit all other $n-1$ agents and assign them rank based on the order they are visited, i.e., the agent visited $i$-th  in the order receives rank $i$. This can again be done by revisiting
the DFS tree $T_{r_l}$ built by $r_v$ while running $DFS(roundNo_v,r_v)$ to elect itself as a leader during Algorithm \ref{algorithm:leader-election} ($Global\_Election$). This revisit finishes in $O(n)$ rounds. 
We use this second method for MST construction. 

As soon as an agent receives its rank, it considers itself as a single component. 
The leader $r_v$ at node $v$ (which has rank-1) starts MST construction. 
$r_v$ includes the MOE adjacent to $v$ in its component and passes a token (message) to the rank-2 agent. This process runs iteratively until the rank-$(n-1)$ agent passes the token to the rank-$n$ agent. Consequently, the rank-$n$ agent includes in its component the MOE available and passes the token to the rank-1 agent ($r_v$). This whole process of passing the token from rank-1 node to rank-$n$ node and back to rank-1 node is one phase. 

In the next phase, the minimum rank agent would include the MOE available to its component and pass the token to the next minimum rank agent, iteratively. In this way, the token reaches the minimum rank agent from the highest rank agent and this phase is completed. This process is repeated phase-by-phase until there is a single component left. Eventually, we have an MST of a single component with $n$ agents. Let us call this algorithm \textit{MST\_Construction}. 
A complete pseudocode is given in Algorithm~\ref{alg: mst_construction}. 

Our algorithm resembles the MST construction algorithm of Gallager, Humblet, and Spira \cite{Gallager83} with the difference that we start MST construction through ranks already provided to agents, whereas in \cite{Gallager83} all nodes have the same rank. The merging of two same rank components with rank $k$ in \cite{Gallager83} provides a new rank of  $k+1$ for the merged component. In ours, there will be no same rank component and hence the merged component gets the rank of one of the components merged.  The token is sent in our algorithm in the order of the component ranks to ensure that all the components can run merging.


\begin{algorithm}[h]
\footnotesize{
\SetKwInput{KwInput}{Input}
\SetKwInput{KwEnsure}{Ensure}
\KwInput{An $n$-node anonymous network with $n$ agents with unique IDs placed on $n$ nodes with an agent $r_l$ at a node elected as a leader (Algorithm \ref{algorithm:leader_election}) with DFS tree $T_{r_l}$  built while running $DFS(roundNo_l,r_l)$ (Algorithm \ref{algorithm:leader-election}).} 
\KwEnsure{MST construction.}    

The leader assumes rank $1$.  It re-traverses the DFS tree $T_{r_l}$ and returns to its home node.  While re-traversing $T_{r_l}$, it provides the distinct rank from the range of $[2, n]$ to the agents at $n-1$ other nodes in order of visit. 
\label{line: rank_numbering}

Each agent $r_u$ considers itself as a component $C_{r_u}$ with $rank(C_{r_u})\leftarrow rank(r_u)$. 
\label{line: single_component}

The leader $r_l$ generates a token.  Let its component be $C_{r_l}$.
 
\While{$|C_{r_l}|<n$}
{   \If{$r_u$ has the token}
    {
        \If{$rank(r_u) \leq rank(C_{r_u})$}
        {
            Agent $r_u$ finds the MOE of the $C_{r_u}$ going to another component $C_{r_w}$ connecting $r_u$ with $r_w$.

            \If{$rank(C_{r_u}) < rank(C_{r_w})$}
            {
                $r_w$ becomes the root node of $C_{r_w}$ by reversing the parent-child pointers from $r_w$ up to the root node of $C_{r_w}$ and $r_u$ becomes the parent of $r_w$ (Figs.~\ref{fig1} and \ref{fig2}). $C_{r_w}$ then merges with $C_{r_u}$ giving a new component $C_{r_u}^{new}$ with $rank(C^{new}_{r_u})\leftarrow rank(C_{r_u})$.
            }
            
            \If{$rank(r_u) < n$}
            {
                $r_u$ passes the token to agent $r_v$ with  $rank(r_v)=rank(r_u)+1$ using $T_{r_l}$.
            }
        }
        \ElseIf{$rank(r_u) > rank(C_{r_u})$ and $rank(r_u) < n$}
        {
            $r_u$ passes the token to agent $r_v$ with  $rank(r_v)=rank(r_u)+1$ using $T_{r_l}$.
        }

        \ElseIf{$rank(r_u) = n$}
        {
        $r_u$ passes the token to the leader $r_l$ (with rank $1$)  using $T_{r_l}$. This token passing visits the parents of $r_u$ until the token reaches the root node of $T_{r_l}$, where the leader is positioned. 
          
        }
    }
} 
\caption{{$MST\_Construction$}}
\label{alg: mst_construction}

        
      
}
\end{algorithm}

\vspace{2mm}
\noindent{\bf Detailed Description of the Algorithm.} 
As discussed earlier, consider that $r_v$ is the leader positioned at node $v$. Leader $r_v$ runs $DFS(r_v)$ to visit all the $non\_candidate$ agents and the $i$-th visited agent receives the rank $i$. The leader position is considered as the first, therefore, the assigned rank for the leader is $1$.  
Let $C_i$ be the component of rank-$i$ agent. Initially, $|C_i|=1$.  We set the rank of component $C_i$ be be the rank of agent $i$, i.e., $rank(C_i)=rank(i)$.
The rank-$1$ agent (leader) generates a token and performs the following step iteratively. The rank-1 agent checks its component size $|C_1|$. If $|C_1|<n$, the leader includes the MOE leading to neighbor agent $r_u$ in $C_1$, and assigns $rank(C_1)\leftarrow rank(C_{r_u})$, where $rank(C_{r_u})$ is the rank of the component $C_{r_u}$ that $r_u$ belongs to.  
The MOE is the edge having one endpoint at a node in $C_i$ ($r_v$'s component) and the other endpoint at a node in $C_j$ ($r_u$'s component). Including the MOE in a component is {\em component merging} which makes two components a single component. 
The leader then passes the token to the next minimum available rank agent $r_w$, which is {\em token passing}. If $rank(C_{r_w})<rank(r_w)$, $r_w$ passes the token to the agent $r_x$ with  $rank(r_x)=rank(r_w)+1$ (without component merging). Otherwise, $r_w$ includes the MOE to its component $C_{r_w}$  and assigns $\forall r\in C_{r_w}, rank(r)\leftarrow rank(C_{r_w})$ 
and pass the token to the agent $r_x$ with  $rank(r_x)=rank(r_w)+1$ This inclusion of MOE is component merging. 
When this is done for all the agents once and rank-$n$ agent passes the token back to the leader, a phase is finished. This process repeats and stops after the leader agent (rank-1) has $|C_1| = n$.


Now, we discuss the \textit{token passing} and \textit{merging} in detail. 

\vspace{1mm}
\noindent\textit{\bf Token Passing:} The tokens passed are of two types:  (1) token passed by the rank-$i$ agent to rank-$(i+1)$ agent (2) token passed by the rank-$n$ agent to rank-$1$ agent (the leader). Both token types are passed using the DFS tree $T_{r_l}$ built by the leader agent $r_v$ while running $DFS(roundNo_v,r_v)$ (during Algorithm \ref{algorithm:leader-election}).   
In the first case, the token passing follows the order in which nodes of $G$ are visited by $DFS(roundNo_v,r_v)$. 
In the second case, the token passed by the rank-$n$  agent follows the parent pointers in $T_{r_l}$ until it reaches the root of $T_{r_l}$ where rank-$1$ agent is positioned.  

\begin{figure}[!t]
\centering
\begin{tikzpicture}[level distance=1.5cm,
  level 1/.style={sibling distance=3cm},
  level 2/.style={sibling distance=1.5cm}]
  \node[circle,draw] (root) {$r_y$}
    child {node[circle,draw] (child1) {$r_x$}
      child {node[circle,draw, red] (child2) {$r_w$}
        child {node[circle,draw] (child8) {$r_h$}} 
      }
      child {node[circle,draw] (child3) {$r_e$}}
    }
    child {node[circle,draw] (child4) {$r_c$}
      child {node[circle,draw] (child5) {$r_f$}}
      child {node[circle,draw] (child6) {$r_g$}}
    };
  \draw[->] (root) -- (child1);
  \draw[->] (child1) -- (child2);
  \draw[->] (child1) -- (child3);
  \draw[->] (root) -- (child4);
  \draw[->] (child4) -- (child5);
  \draw[->] (child4) -- (child6);
  \draw[->] (child2) -- (child8);
  
  \begin{scope}[xshift=-4.5cm]
  \node[circle,draw] (root2) {$r_a$}
    child {node[circle,draw, red] (child42) {$r_u$}
      child {node[circle,draw] (child52) {$r_v$}}
      child {node[circle,draw] (child62) {$r_z$}}
    };
  \draw[->] (root2) -- (child42);
  \draw[->] (child42) -- (child52);
  \draw[->] (child42) -- (child62);
\draw[dotted, blue] (child42) -- (child2) node[midway, above, font=\footnotesize, blue] {MOE};
  \end{scope}
\end{tikzpicture}
\caption{The components $C_{r_w}$ and $C_{r_u}$ before $C_{r_w}$ merges with $C_{r_u}$ due to the MOE connecting node $r_u\in C_{r_u}$ with node  $r_w\in C_{r_w}$ and $rank(C_{r_u})<rank(C_{r_w})$ (if $rank(C_{r_w})<rank(C_{r_u})$, then $C_{r_u}$ merges with $C_{r_w}$ due to the MOE connecting $r_w\in C_{r_w}$ with node  $r_u\in C_{r_u}$). Both $C_{r_u}$ and $C_{r_w}$ are rooted trees with roots $r_a$ and $r_y$, respectively. In the figure, $r_A \rightarrow r_B$ denotes $r_A$ is the parent of $r_B$.} 
\label{fig1}
\vspace{-3mm}
\end{figure}
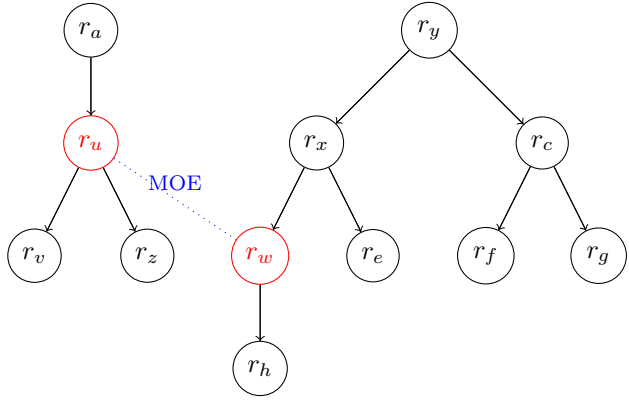

\begin{figure}
\centering
\begin{tikzpicture}[level distance=1.5cm,
  level 1/.style={sibling distance=3cm},
  level 2/.style={sibling distance=1.5cm}]
  \node[circle,draw] (root) {$r_y$}
    child {node[circle,draw] (child1) {$r_x$}
      child {node[circle,draw, red] (child2) {$r_w$}
        child {node[circle,draw] (child8) {$r_h$}} 
      }
      child {node[circle,draw] (child3) {$r_e$}}
    }
    child {node[circle,draw] (child4) {$r_c$}
      child {node[circle,draw] (child5) {$r_f$}}
      child {node[circle,draw] (child6) {$r_g$}}
    };
  \draw[<-,red] (root) -- (child1);
  \draw[<-, red] (child1) -- (child2);
  \draw[->] (child1) -- (child3);
  \draw[->] (root) -- (child4);
  \draw[->] (child4) -- (child5);
  \draw[->] (child4) -- (child6);
  \draw[->] (child2) -- (child8);
  
  \begin{scope}[xshift=-4.5cm]
  \node[circle,draw] (root2) {$r_a$}
    child {node[circle,draw, red] (child42) {$r_u$}
      child {node[circle,draw] (child52) {$r_v$}}
      child {node[circle,draw] (child62) {$r_z$}}
    };
  \draw[->] (root2) -- (child42);
  \draw[->] (child42) -- (child52);
  \draw[->] (child42) -- (child62);
\draw[->, red] (child42) -- (child2) node[midway, above = 0.1cm, font=\footnotesize, blue] {MOE};
  \end{scope}
\end{tikzpicture}
\caption{The resulting component $C^{new}_{r_u}$ after merging $C_{r_w}$ with $C_{r_u}$. Component $C^{new}_{r_u}$ gets rank $rank(C^{new}_{r_u})\leftarrow rank(C_{r_u})$ since $C_{r_w}$ merged with $C_{r_u}$ to become $C^{new}_{r_u}$ due to $rank(C_{r_u})<rank(C_{r_w})$. Two parent-child pointers in  $C_{r_w}$ are reversed to keep $C^{new}_{r_u}$ a rooted tree. 
}
\label{fig2}
\vspace{-3mm}
\end{figure}
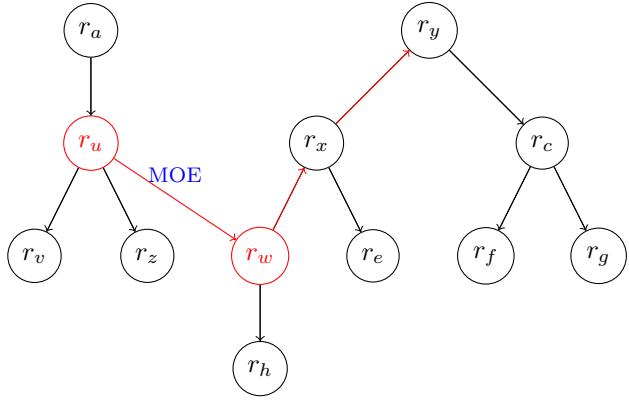

\vspace{1mm}
\noindent\textit{\bf Merging:}\label{merging} 
Suppose agent $r_u$ with $1\leq rank(r_u)\leq n$ in component $C_{r_u}$ has the token and $rank(r_u)\leq rank(C_{r_u})$. $C_{r_u}$ is a tree and has a root node/agent. 
Agent $r_u$ finds the  MOE connected to $C_{r_u}$ (the component it belongs to) as follows. 
Agent $r_u$ traverses 
$C_{r_u}$ and checks the incident edges of the nodes in $C_{r_u}$ (edges with both endpoints on the nodes in $C_{r_u}$ are not considered)  in the ascending order of the  (edge) weights. 
$r_u$ adds the MOE among the incident edges to $C_{r_u}$. Let the (added) MOE have the other end at a node $w$ that belongs to $C_{r_w}$. W.l.o.g., let us consider $rank(C_{r_u}) < rank(C_{r_w})$ (otherwise, $r_u$ simply passes the token to agent with $rank(r_u)+1$).
The old parent, say $r_x$ (from the component $C_{r_w}$), of $r_w$ becomes $r_w$'s child. The process of converting a parent to a child starting from $r_w$ runs until reaching the root node in $C_{r_w}$. Each subsequent parent now becomes a child and the child becomes a parent, i.e.,  $r_w$ becomes the root of $C_{r_w}$. Since MOE is added, $r_u$ becomes the parent of $r_w$.  $C_{r_u}$ and $C_{r_w}$ now become a new single component $C^{new}_{r_u}$. Since $rank(C_{r_u})<rank(C_{r_w})$, $rank(C^{new}_{r_u})\leftarrow rank(C_{r_u})$, which is communicated to all the agents (nodes) in $C^{new}_{r_u}$. 
Furthermore, the $C_{r_u}$'s component does not go through any transition with respect to parent-child pointers except the fact that $r_u$ becomes the parent of $r_w$. 
Figs.~\ref{fig1} and \ref{fig2} illustrate these ideas. Fig.~\ref{fig1} shows components $C_{r_u}$ and $C_{r_w}$ before they merge due to the MOE connecting $r_u$ with $r_w$. Fig.~\ref{fig2} captures the merged component $C^{new}_{r_u}$ such that  root of the  $C^{new}_{r_u}$ remains unchanged and the pointer changes occurred in the $C_{r_w}$ component during the merging. The directed edge denotes new parent-child relationships.

\vspace{2mm}
\noindent{\bf Analysis of the Algorithm.}
We now analyze Algorithm \ref{alg: mst_construction} for its correctness, time, and memory complexities. The correctness proof shows Algorithm \ref{alg: mst_construction} indeed constructs an MST.

\begin{lemma}
    Algorithm~\ref{alg: mst_construction} 
    generates the MST of $G$. \label{lem: mst_correctness}
\end{lemma}
\begin{proof}
We prove this in three steps: firstly, Algorithm~\ref{alg: mst_construction} constructs a tree; secondly,  the constructed tree is a spanning tree; finally, the spanning tree is, indeed, a minimum spanning tree. Firstly, we prove by contradiction that no cycle is generated during Algorithm~\ref{alg: mst_construction}. Let us suppose, there exists a cycle at any point during the algorithm. Then it implies two components with the same rank merged at some point, which is a contradiction. Secondly, let us suppose there exist at least two components at the end of the algorithm. This implies that the leader component (rank-$1$) did not merge with the other component and terminated the algorithm, which contradicts the fact that the algorithm terminates when the leader is connected to $n$ agents altogether.

Finally, consider that the tree formed by our algorithm is $T$ and the MST is $T^{*}$. Note that in Algorithm~\ref{alg: mst_construction}, each edge added to the MST tree is by selection of a MOE.  If $T = T^{*}$ then $T$ is minimum spanning tree. If $T \neq T^{*}$ then there exists an edge $e \in T^{*}$ of minimum weight such that $e \notin T$. Therefore, there exists a phase in which $e$ was not considered during component merging, and an edge with higher weight, say $e'$, was considered. But this is contradictory to Algorithm~\ref{alg: mst_construction} which merges the components with a MOE. Therefore, Algorithm~\ref{alg: mst_construction} constructs the MST of $G$.
\end{proof}



\begin{lemma}\label{lem: phase_number}
    In Algorithm~\ref{alg: mst_construction}, 
    the leader initiates the merging $O(\log n)$ times.
\end{lemma}
\begin{proof}
    Initially, there are $n$ single-node components in Algorithm~\ref{alg: mst_construction} (Line~\ref{line: single_component}). In each phase (leader initiating a token until the token returns to the leader), each component merges at least once with another component. Therefore, after every phase, the number of components is reduced by at least half. Consequently, after $O(\log n)$ phases, there remains only a single component of $n$ nodes. 
\end{proof}


\noindent{\bf Proof of Theorem \ref{theorem:result}:}
\begin{proof}
Providing ranks to agents takes $O(n)$ rounds by re-traversing the DFS tree $T_{DFS}$ constructed by $DFS(roundNo_u,r_u)$ during leader election (Algorithm \ref{algorithm:leader-election}). The while loop performs two operations - token passing and merging.

In token passing, the token is passed through the edge of the tree $T_{DFS}$, and an edge is not traversed more than twice. Therefore, in a phase, to pass the token from rank-$1$ agent to rank-$n$ agent takes $O(n)$ rounds. From rank-$n$ agent the token returns to the leader again in $O(n)$ rounds. Combining this with Lemma~\ref{lem: phase_number}, token passing takes $O(n \log n)$ rounds. 

In the process of merging, an agent $r_u$ visits at most three types of edges: i) MOE edges within its component $C_{r_u}$ ii) edges traversed to find MOE iii) reversing the edges from $r_w$ until the root of $C_{r_w}$ when it merges with another component $C_{r_u}$ at $r_w$. In the case of i) MOE is the part of the component $C_{r_u}$, i.e., a tree. Its traversal finishes in $O(|C_{r_u}|)$ rounds. In a phase, the combined size of all the components is $O(n)$. In case ii) edges that are traversed to find the MOE were either part of MOE or not, in case, they become part of MOE they were traversed two times. There are at most $(n-1)$ such edges throughout the process. On the other hand, if some edges did not become part of the MOE then they were never traversed again. Therefore, there are in total $m-(n-1)$ such edges. In case iii) reversing an edge from its merging point to the root can not be more than its component size. Therefore, reversing of edge for agent $r_u$ takes $O(|C_{r_u}|)$. Combining the time for the cases i) and iii) per phase with $O(\log n)$ phases (Lemma~\ref{lem: phase_number}), we have total runtime  $O(n \log n)$ rounds and for case ii) we have total $O(m)$ rounds throughout the execution. Thus, the overall round complexity becomes $O(m + n \log n)$.


For memory, rank numbering takes  $O(\log n)$ bits at each agent to re-traverse the DFS tree $T_{DFS}$ (constructed during Algorithm \ref{algorithm:leader-election}). Furthermore, each agent stores $O(\log n)$ bits to keep the account of the ID/rank and component rank. Also, there might be a case in which all the neighbors are part of the MST. Therefore, in the worst case, the highest degree ($\Delta$) agent (agent placed at the highest degree node) keeps the account of all the MST edges and requires $O(\Delta \log n)$ memory. Hence, the overall memory required by each agent in Algorithm~\ref{alg: mst_construction} is $O(\Delta \log n)$ bits. Combining the memory needed for leader election: (i) $O(\log^2n)$ bits per agent (for dispersed initial configuration) and (ii) $O(n\log n)$ bits per agent  (for rooted and general initial configurations), we have the theorem.
\end{proof}

\section{Applications to Other Graph Problems}
\label{section:applications}

We apply the leader election result (Theorem \ref{theorem:leader}) and improve the existing results on Gathering, MIS, and MDS problems in the agent-based model. The previous results on these problems in the agent-based model considered some graph parameters known to agents a priori (Table \ref{table:results}). We lifted those assumptions and additionally improved the time/memory complexities.

\vspace{2mm}
\noindent{\bf  Gathering.} Suppose $n$ agents are initially located arbitrarily on the nodes of an $n$-node anonymous graph $G = (V,E)$. The goal of the gathering problem is to relocate the agents autonomously to position all of them to a node in $G$ not fixed a priori ($n$ agents at a node of $G$). Recently, Molla  {\it et al.} \cite{Molla-2021IPDPS} established the following theorem in the agent-based model. 

\begin{theorem}[Molla {\it et al.} \cite{Molla-2021IPDPS}]
\label{theorem:Gathering_previous}
Given any configuration of $n$ agents positioned initially arbitrarily on the nodes of a $n$-node graph $G$, there is a deterministic algorithm that collects all $n$ agents to a node not fixed a priori in $G$ in $O(n^3)$ rounds with $O(M+m\log n)$ bits at each agent, with agents knowing $n$ a priori, where $M$ is the memory required to implement the Universal Exploration Sequence (UXS) procedure \cite{Ta-ShmaZ14}.  \end{theorem}

Using our result on leader election, we establish the following result. 

\begin{theorem}
\label{theorem:Gathering}
Given any configuration of $n$ agents positioned initially arbitrarily on the nodes of a $n$-node graph $G$, there is a deterministic algorithm that collects all $n$ agents to a node in $G$ not fixed a priori in $O(m)$ rounds with $O(n \log n)$ bits at each agent for general initial configurations and with $O(\log^2n)$ bits at each agent for dispersed initial configurations, without agents knowing any graph parameter a priori.  
\end{theorem}

Theorem \ref{theorem:Gathering} is an improvement to Theorem \ref{theorem:Gathering_previous} w.r.t. three aspects: (i) since $m\leq n^2$, at least an $O(n)$ factor is removed from time complexity, (ii) the memory per agent is improved from $O(M+m\log n)$ bits to $O(n\log n)$ bits ($O(\log^2n)$ bits for dispersed initial configurations), and (iii) a priori knowledge on $n$ is lifted. 

Our idea in establishing Theorem \ref{theorem:Gathering} is as follows. We first elect a leader using Algorithm \ref{algorithm:leader_election} which finishes in $O(m)$ rounds with $O(n\log n)$ bits at each agent for general initial configurations and $O(\log^2n)$ bits at each agent for dispersed initial configurations. We then ask the leader to re-traverse the DFS tree built during the leader election collecting the agents settled at the graph nodes which finishes in $O(n)$ time 
with $O(n\log n)$ bits per agent (for general initial configurations) and with $O(\log^2n)$ bits per agent (for dispersed).

\vspace{2mm}
\noindent{\bf  MIS.} 
Suppose $n$ agents are initially located arbitrarily on the nodes of an $n$-node anonymous graph $G = (V,E)$. The goal in the maximal independent set (MIS) problem is to relocate the agents autonomously to find a subset $S \subset V$ of nodes such that $S$ forms an MIS of $G$. Recently, Pattanayak {\it et al.} \cite{PattanayakBCM24} established the following theorem. 

\begin{theorem}[Pattanayak {\it et al.} \cite{PattanayakBCM24}]
\label{theorem:MIS_previous}
Given any configuration of $n$ agents positioned initially arbitrarily on the nodes of a $n$-node graph $G$, there is a deterministic algorithm that finds an MIS of $G$ in $O(n\Delta \log n)$ rounds with $O(\log n)$ bits at each agent, if agents know graph parameters $n$ and $\Delta$ a priori.   
\end{theorem}

Using our result on leader election, we establish the following result. 

\begin{theorem}
\label{theorem:MIS}
Given any configuration of $n$ agents positioned initially arbitrarily on the nodes of a $n$-node graph $G$, there is a deterministic algorithm that finds an MIS of $G$ in $O(n\Delta)$ rounds with (i) $O(n\log n)$ bits at each agent starting from rooted or general initial configurations and (ii) $O(\log^2n)$ bits per agent starting from dispersed initial configurations, without agents knowing  $n$ and $\Delta$ a priori.   
\end{theorem}

Theorem \ref{theorem:MIS} improves Theorem \ref{theorem:MIS_previous} in two aspects: (i) an $O(\log n)$ factor is removed from time complexity and (ii) the assumption of a priori knowledge of $n$ and $\Delta$ is lifted.

Our idea in establishing Theorem \ref{theorem:MIS} is as follows. We first elect a leader using Algorithm \ref{algorithm:leader_election} which finishes in $O(m)$ rounds with $O(n\log n)$ bits at each agent (or $O(\log^2n)$ bits per agent for dispersed initial configurations). We then gather all $n$ agents 
at the leader node, which finishes in $O(m)$ rounds with $O(n\log n)$ bits at each agent ($O(\log^2n)$ for dispersed). We then construct an MIS using the technique of  Pattanayak {\it et al.} \cite{PattanayakBCM24} which finishes finding an MIS in $O(n\Delta)$ rounds with $O(\log n)$ bits at each agent. 
Therefore, the runtime complexity becomes $O(n\Delta)$ rounds, since $m\leq n\Delta$, and memory remains $O(n\log n)$ bits per agent for rooted and general initial configurations and $O(\log^2n)$ bits for dispersed initial configurations.  

\vspace{2mm}
\noindent{\bf  Minimal Dominating Sets.}
Suppose $n$ agents are initially located arbitrarily on the nodes of an $n$-node anonymous graph $G = (V,E)$. 
A {\em dominating set} of $G$ 
 is a subset $DS\subset V$ of nodes such that for any 
$v\notin DS$, $v$ has a neighbour in $DS$. Recently, Chand {\it et al.} \cite{ChandMS23} established the following theorem. 

\begin{theorem}[Chand {\it et al.} \cite{ChandMS23}]
\label{theorem:MDS_previous}
Given any configuration of $n$ agents positioned initially arbitrarily on the nodes of a $n$-node graph $G$, there is a deterministic algorithm that finds a minimal dominating set (MDS) of $G$ in $O(\gamma \Delta \log n+n\gamma+m)$ rounds with $O(\log n)$ bits at each agent, if agents know graph parameters $n,\Delta,m,\gamma$ a priori, where $\gamma$ is the number of clusters of agents in the initial configuration. 
\end{theorem}

Using our result on leader election, we establish the following result. 

\begin{theorem}
\label{theorem:MDS}
Given any configuration of $n$ agents positioned initially arbitrarily on the nodes of a $n$-node graph $G$, there is a deterministic algorithm that finds an MDS of $G$ in $O(m)$ rounds with (i) $O(n\log n)$ bits at each agent starting from rooted or general initial configurations and (ii) $O(\log^2n)$ bits per agent starting from dispersed initial configurations, without agents knowing $n,\Delta,m,\gamma$ a priori.   
\end{theorem}

Theorem \ref{theorem:MDS} improves Theorem \ref{theorem:MDS_previous} in two aspects: (i) an $O(\gamma \Delta \log n+n\gamma)$ additive factor is removed from time complexity and (ii) a priori knowledge on $n,\Delta,m,\gamma$ is lifted.

Our idea in establishing Theorem \ref{theorem:MDS} is as follows. We first elect a leader using Algorithm \ref{algorithm:leader_election} which finishes in $O(m)$ rounds with $O(n\log n)$ bits at each agent (or $O(\log^2n)$ bits per agent for dispersed initial configurations). We then gather all $n$ agents at the leader node, which finishes in $O(m)$ rounds with $O(n\log n)$ bits at each agent ($O(\log^2n)$ bits if dispersed initially). We then construct an MDS of $G$ using the technique of  Chand {\it et al.} \cite{ChandMS23} which finishes in $O(m)$ rounds with $O(\log n)$ bits at each agent starting from the rooted configuration. 
Therefore, the runtime becomes $O(m)$ and memory remains $O(n\log n)$ bits per agent for rooted and general initial configurations and $O(\log^2n)$ bits for dispersed initial configurations.

\section{Concluding Remarks}
\label{section:conclusion}
\noindent{\bf Discussion on Memory Requirement.} In our leader election algorithm, if $n$ and $\Delta$ are known, a dispersed configuration can be achieved starting from any initial configuration in either $O(n\log^2n)$ rounds using the algorithm of Sudo {\it et al.} \cite{sudo23} or in $O(m)$ rounds using the algorithm of Kshemkalyani and Sharma \cite{KshemkalyaniS21-OPODIS}, with $O(\log n)$ bits per agent. After that, the singleton election procedure can finish in $O(\Delta\log^2n)$ rounds with $O(\log n)$ bits per agent. Then finally the local leaders can run a global election procedure to elect a unique global leader in $O(m)$ rounds with $O(\log n)$ bits per agent. Therefore, leader election only needs $O(\log n)$ bits per agent ($n$ factor improvement compared to our algorithm for rooted and general initial configurations and $\log n$ factor improvement for dispersed initial configurations). 
For the MST construction, a node may need to remember multiple of its neighboring edges as a part of MST and hence the total memory needed is $O(\Delta\log n)$ bits per agent, which also includes the memory per agent for leader election. 
However, notice that this memory improvement is achieved knowing $n$ and $\Delta$. The proposed leader election algorithm achieves the claimed bounds (Theorem \ref{theorem:leader}) without knowing any graph parameter a priori.
This result helped to achieve for the first time results for MST in the agent-based model and also to provide improved time/memory results for gathering, MIS, and MDS in the agent-based model, lifting the assumptions on known graph parameters.

\vspace{1mm}
\noindent{\bf Concluding Remarks.}
We have initiated the study of the leader election in the agent-based model. The considered agent-based model poses unique challenges compared to the well-studied message-passing model. We have developed four results. The first result provides a deterministic solution in the agent-based model simulating an existing deterministic algorithm in the message-passing model under assumptions of known graph parameters $n$ and $\Delta$ and with memory at each agent proportional to the node memory in the message-passing model. The second result elects a leader in the agent-based model, with $n$ agents starting arbitrarily initially, in $O(m)$ time with $O(n\log n)$ bits per agent for rooted and general initial configurations and $O(\log^2n)$ bits for dispersed initial configurations, without agents knowing any graph parameter a priori.
The third result constructs an MST of $G$, given an elected leader, in $O(m+n\log n)$ time with $O(\Delta \log n)$ bits at each agent, without agents knowing any graph parameter a priori. 
Finally, the time complexities of the existing results on Gathering, MIS, and MDS on the agent-based model were improved using the leader election result, removing the assumptions on agents knowing graph parameters a priori.   

\vspace{1mm}
\noindent{\bf Future Work.} It would be interesting to improve the time/memory complexities of our leader election and MST solutions, 
without agents knowing graph parameters a priori.  It would also be interesting to solve other fundamental distributed graph problems, such as coloring, maximal matching, and minimum cut, in the agent-based model. 

\bibliography{references}


\end{document}